\documentclass[10pt,conference]{IEEEtran}
\IEEEoverridecommandlockouts

\usepackage{amsmath,amsfonts,amsthm,amssymb}
\usepackage{algorithm}
\usepackage{algorithmicx}
\usepackage{algpseudocode}
\usepackage{graphicx}
\usepackage{textcomp}
\usepackage[usenames,dvipsnames,table]{xcolor}
\usepackage[hidelinks]{hyperref}
\usepackage[numbers,sort&compress]{natbib}
\usepackage{colortbl}
\usepackage[capitalize]{cleveref}
\usepackage{xurl}
\usepackage{enumitem}
\usepackage{booktabs}
\usepackage{multirow}
\usepackage{tabularx}
\usepackage{tikz}
\usepackage[most]{tcolorbox}
\usepackage{pifont}
\usepackage{latexsym}
\usepackage[T1]{fontenc}
\usepackage[utf8]{inputenc}
\usepackage{microtype}
\usepackage{xspace}
\usepackage{fontawesome5}
\usepackage{wrapfig}
\usepackage{makecell}
\usepackage{bbm}
\usepackage{hhline}
\usepackage{placeins}
\usepackage{tablefootnote}
\usepackage{graphicx}

\Crefname{section}{Sec.}{Sects.}
\setitemize{noitemsep,topsep=0pt,parsep=0pt,partopsep=0pt}

\def\BibTeX{{\rm B\kern-.05em{\sc i\kern-.025em b}\kern-.08em
    T\kern-.1667em\lower.7ex\hbox{E}\kern-.125emX}}

\definecolor{headercolor}{RGB}{248, 196, 145}
\definecolor{rowlight}{RGB}{255, 243, 230}
\definecolor{rowwhite}{RGB}{255, 255, 255}
\definecolor{bestred}{RGB}{210, 45, 45}
\definecolor{secondblue}{RGB}{40, 100, 195}
\definecolor{rulecolor}{RGB}{240, 200, 160}

\definecolor{blue(ncs)}{rgb}{0.0, 0.53, 0.74}

\definecolor{realgreen}{RGB}{34, 120, 79}
\definecolor{simbrown}{RGB}{160, 82, 45}
\definecolor{tagblue}{RGB}{52, 101, 175}
\definecolor{taggray}{RGB}{130, 130, 130}
\definecolor{tagorange}{RGB}{200, 120, 20}
\definecolor{tagred}{RGB}{190, 50, 50}
\definecolor{missred}{RGB}{220, 60, 60}

\tikzset{
  tag/.style={rounded corners=2pt, inner sep=1.5pt, outer sep=0pt,
              font=\sffamily\bfseries\tiny, minimum width=0.8cm, text=white, align=center},
  tag edit/.style={tag, fill=tagblue},
  tag sel/.style={tag, fill=taggray},
  tag cpl/.style={tag, fill=tagorange},
  tag term/.style={tag, fill=realgreen},
  tag err/.style={tag, fill=tagred},
  tag chat/.style={tag, fill=tagorange!80!red},
  op desc/.style={font=\tiny, anchor=west, text width=3.5cm},
  timeline/.style={line width=1.2pt},
  node dot/.style={circle, inner sep=1.2pt, fill=#1},
}

\definecolor{listCircle}{RGB}{100, 181, 226}

\newcounter{resq}

\crefname{resq}{RQ}{RQs}
\crefformat{resq}{RQ#2#1#3}
\Crefname{resq}{RQ}{RQs}

\newcommand{\challengename}{Challenge}
\newcounter{challenge}
\crefformat{challenge}{#2\challengename\ #1#3}
\Crefformat{challenge}{#2\challengename\ #1#3}

\algrenewcommand\algorithmicrequire{\textbf{Input:}}
\algrenewcommand\algorithmicensure{\textbf{Output:}}

\def\name{\textit{\textsc{MEMCoder}}\xspace}

\def\ndonnx{{\fontfamily{cmtt}\selectfont{NdonnxEval}}\xspace}
\def\numba{{\fontfamily{cmtt}\selectfont{NumbaEval}}\xspace}

\newcommand{\bi}[1]{\textbf{\textit{#1}}}
\newcommand{\pass}[1]{\textit{pass@#1}}
\newcommand{\exec}[1]{\textit{exec@#1}}

\newcommand{\baselinename}[1]{\textit{#1}\xspace}
\newcommand{\llmname}[1]{{\fontfamily{pcr}\selectfont {#1}}\xspace}

\definecolor{namerow}{RGB}{246,249,255}
\definecolor{vanillarow}{RGB}{252,252,252}
\definecolor{ragrow}{RGB}{246,252,255}
\definecolor{newnamerow}{RGB}{246,253,246}

\newtcolorbox{boxK}{
    top=2pt,
    bottom=2pt,
    left=3pt,
    right=3pt,
    boxrule = 0pt,
    toprule = 0pt,
    enhanced,
    fuzzy shadow = {0pt}{-1pt}{-0.2pt}{0.2pt}{black!35}
}

\begin{document}

\title{Learning from Execution: Self-Evolving Memory for Private-Library Code Generation
}

\author{\IEEEauthorblockN{Mofei Li}
\IEEEauthorblockA{\textit{College of AI} \\
\textit{Tsinghua University}\\
Beijing, China \\
\textit{Fitten Tech Co., Ltd.}\\
Beijing, China \\
lmf25@mails.tsinghua.edu.cn}
\and
\IEEEauthorblockN{Taozhi Chen}
\IEEEauthorblockA{\textit{College of AI} \\
\textit{Tsinghua University}\\
Beijing, China \\
\textit{Emory University}\\
Atlanta, GA, USA \\
taozhi.chen@emory.edu}
\and
\IEEEauthorblockN{Guowei Yang}
\IEEEauthorblockA{
\textit{Fitten Tech Co., Ltd.}\\
Beijing, China \\
gavin@eniacode.com}
\and
\IEEEauthorblockN{Jia Li\textsuperscript{$\dagger$}}
\IEEEauthorblockA{\textit{College of AI} \\
\textit{Tsinghua University}\\
Beijing, China \\
jia\_li@mail.tsinghua.edu.cn}
}

\maketitle

\begingroup
\renewcommand\thefootnote{}
\footnotetext{\textsuperscript{$\dagger$} Corresponding author.}
\endgroup

\begin{abstract}\looseness=-1
Large Language Models (LLMs) have achieved strong performance on general code generation, but their effectiveness drops sharply in enterprise settings where software development relies on internal private libraries absent from public pre-training corpora. Existing Retrieval-Augmented Generation (RAG) methods provide a training-free solution by retrieving static API documentation, but our analysis shows that documentation mainly helps models identify \textit{what APIs to use} and remains insufficient for teaching \textit{how to use them correctly}. Even with oracle API-document retrieval, LLMs still make recurring errors at the API, cross-API, and task levels, including API misuse or hallucination, flawed API composition, and incorrect solution strategies. To address this limitation, we propose \name, a training-free self-evolving memory framework for private-library code generation. \name augments existing RAG pipelines with a \textbf{Multi-level Evolving Memory} that continuously accumulates and reuses execution-derived \textit{Usage Guidelines} at the API, cross-API, and task levels. During generation, \name retrieves both static API documentation and relevant historical memories to guide code generation; after execution, it analyzes feedback to refine memory through a closed loop of generation, execution, reflection, and update. Extensive experiments on \ndonnx and \numba show that \name consistently enhances different RAG backbones across LLMs of different scales, yielding an average absolute \pass{1} improvement of \textbf{18.41 percentage points}. Moreover, \name outperforms existing self-evolving memory methods and validates the effectiveness of organizing execution feedback into multi-level usage memories.
\end{abstract}

	

\section{Introduction}
\label{sec:introduction}

\looseness=-1
Large Language Models (LLMs) have demonstrated exceptional proficiency in general code generation tasks~\cite{li2025structured, jiang2025aixcoder, li2025beyond, cai2025ai}. However, their performance drops sharply in real-world enterprise environments, where software development often relies on internal private libraries whose APIs, documentation, and usage examples are largely absent from public pre-training corpora~\cite{exploracoder, apifinder}. This gap defines the task of \textit{Private-Library Code Generation}, where models must generate correct code by understanding and invoking APIs from previously unseen private libraries. Since LLMs typically lack prior knowledge of these libraries, they struggle to use them correctly and effectively, significantly limiting their practical utility in real-world software development.


\looseness=-1
To address the issue that private-library APIs are absent from public pre-training corpora, existing methods commonly adopt Retrieval-Augmented Generation (RAG) to provide LLMs with relevant static API documentation at generation time~\cite{apifinder, docprompting, epigen}. However, our analysis in Section~\ref{sec:motivation} shows that RAG mainly helps LLMs identify \textit{what APIs to use}, but provides limited support for \textit{how to use them correctly}. Even under an Oracle setting, where all required API documents are provided, LLMs still produce many failed solutions. Further case studies show that these failures appear at three levels. At the API level, LLMs may incorrectly invoke existing APIs or generate hallucinated APIs that do not exist in the documentation. At the cross-API level, LLMs may retrieve relevant APIs but fail to correctly handle the input-output relationships and composition patterns among multiple APIs. At the task level, LLMs may choose incorrect overall problem-solving steps, causing the final implementation to fail. More importantly, these errors do not occur in isolation. The same type of error may recur across multiple generations of the same task or reappear in different tasks. The root cause lies in the fact that existing RAG methods merely allow LLMs to temporarily read static documentation during each generation; as a result, each generation is similar to encountering the private library from scratch, leaving LLMs persistently unfamiliar with how the library should be used.

A key feature of code generation is that generated code can be executed and verified, allowing explicit feedback signals to be obtained from each generation, such as test outcomes, runtime errors, and assertion failures. These feedback signals naturally contain valuable information about how private libraries are used. Successful code provides executable examples of API invocations, API coordination, and task-solving strategies, while failed executions reveal where LLMs tend to make mistakes when using private libraries and can provide guidance for avoiding similar errors in subsequent generations. In other words, generated code and its execution feedback can serve as a source of \textit{Usage Guidelines} regarding \textit{how to use} private libraries. To make these guidelines reusable in subsequent tasks rather than lost after a single generation, an external memory is needed to store, retrieve, and update them. As the memory continuously evolves within the closed-loop cycle of generation--execution--feedback--update, LLMs can gradually develop familiarity with how to use private libraries.

\begin{figure}[t]
    \centering
    \setlength{\abovecaptionskip}{0pt}
    \setlength{\belowcaptionskip}{0pt}
    \includegraphics[
        width=0.78\columnwidth,
        trim={1cm 1.5cm 9cm 1cm},
        clip
    ]{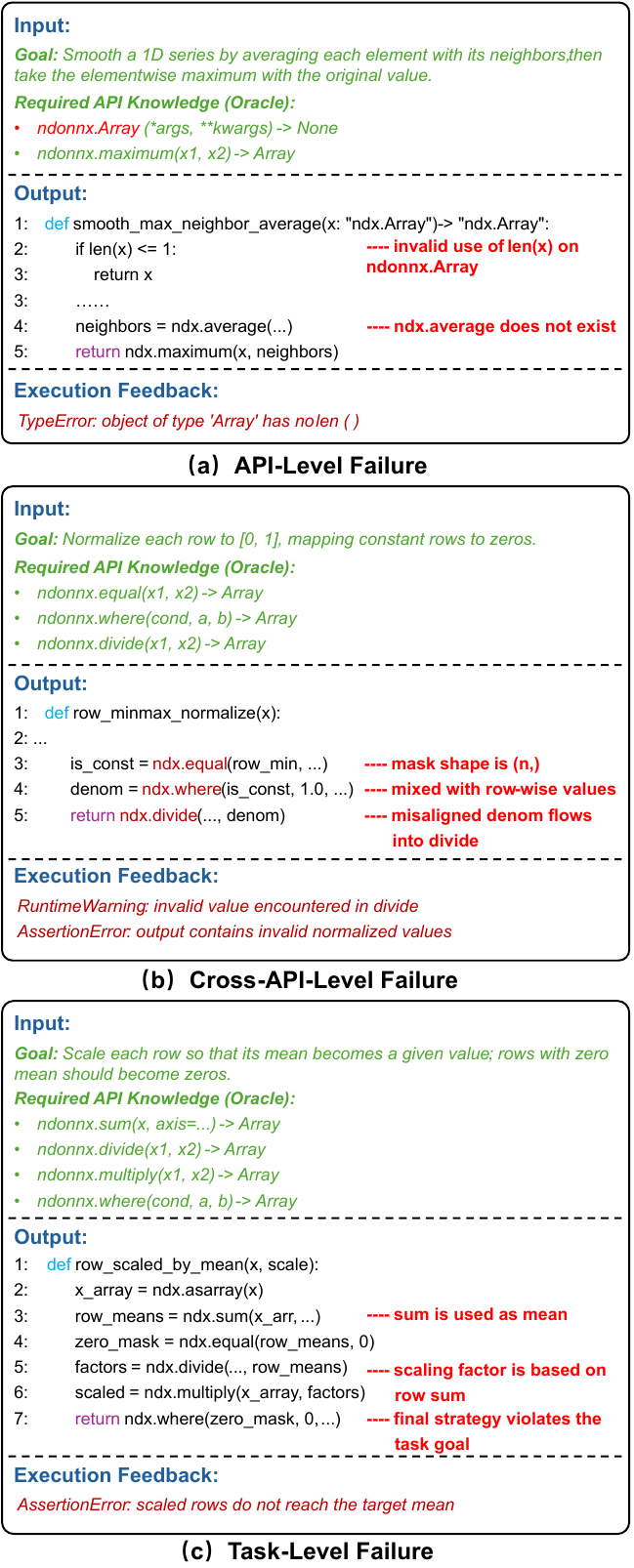}
    \vspace{-4pt}
    \caption{Representative failures under oracle API documentation, showing API-level, Cross-API-level, and Task-level errors that cannot be resolved by static API documents alone.}
    \vspace{-14pt}
    \label{fig:cases}
\end{figure}

\looseness=-1
Based on the above observations and analysis, we propose \name, a training-free self-evolving memory framework for private-library code generation. As an enhancement to existing RAG workflows, \name improves private-library code generation by continuously accumulating and reusing \textit{Usage Guidelines} derived from code execution feedback. Specifically, \name constructs a \textbf{Multi-level Evolving Memory} that maintains API-level, cross-API-level, and task-level memories to capture usage patterns of individual APIs, collaboration patterns among multiple APIs, and solution strategies for similar tasks. During generation, \name first relies on the underlying RAG method to retrieve static API documentation, and then further retrieves relevant memories from the three levels to guide code generation. After execution, \name updates the memory according to the generated code and its execution feedback, allowing the memory to continuously evolve through use. We conduct experiments on two private-library code generation benchmarks across LLMs of different scales, and the results show that \name consistently enhances different RAG methods, achieving an average absolute \pass{1} improvement of 18.41 percentage points, while outperforming general self-evolving methods adapted to this setting.

\looseness=-1
In summary, our contributions are as follows:
\begin{itemize}
    \item We reveal a key limitation of existing RAG methods for private-library code generation: static API documentation mainly helps identify \textit{what APIs to use}, but struggles to support \textit{how to use them correctly}, leading to failures at the API, cross-API, and task levels.

    \item We propose \name, a training-free self-evolving memory framework for private-library code generation, which continuously accumulates and reuses \textit{Usage Guidelines} from execution feedback through API-level, cross-API-level, and task-level memories.

    \item We evaluate \name on two private-library code generation benchmarks across LLMs of different scales. The results show that \name consistently enhances different RAG methods and outperforms general-purpose self-evolving baselines adapted to this task.
\end{itemize}

\section{Related Work}
\label{sec:related-work}



\subsection{Private-Library Code Generation}
\looseness=-1
Private-library code generation requires LLMs to solve programming tasks with APIs that are rarely covered by public training corpora~\cite{apifinder}. Unlike general code generation~\cite{zan2024diffcoder, gu2025effectiveness, liu2025think, liu2023codegen4libs}, LLMs usually have little prior knowledge of the target private libraries, while continual pre-training or fine-tuning is often impractical due to limited supervision data and evolving APIs~\cite{zhang2026masterteachingllmsuse, zan2025private, li2024revisiting, luo2025empirical}. Therefore, existing training-free methods commonly adopt retrieval-augmented generation (RAG)~\cite{lewis2020retrieval}, retrieving relevant APIs or library documents and injecting them into the prompt, such as APIFinder~\cite{apifinder} and DocPrompting~\cite{docprompting}. Later methods further improve API retrieval and context organization by decomposing complex requirements into subtasks to better match and utilize relevant API documentation~\cite{epigen, capir, exploracoder}. Overall, these methods mainly focus on helping LLMs access more accurate or complete API documentation at inference time. However, private-library code generation requires not only knowing which APIs to use, but also understanding API-specific usage constraints, cross-API coordination patterns, and task-level solution strategies. In contrast, \name converts execution feedback into reusable \textit{Usage Guidelines} and continuously retrieves and updates them through a self-evolving memory.

\subsection{Self-Evolving Memory}
\looseness=-1
Recent work has explored how language models can improve over time by accumulating and reusing memory during inference without updating model parameters~\cite{suzgun2025dynamic, wei2025evomem}. These methods typically augment the model with dynamically updated external memory, such as retrieved contexts, task summaries, reflections, execution trajectories, or other experience records, and reuse them in subsequent tasks. Representative approaches, such as Dynamic Cheatsheet~\cite{suzgun2025dynamic} and ReMem~\cite{wei2025evomem}, show that inference-time experience accumulation can improve downstream performance. However, existing methods are mainly designed for mathematical reasoning, general problem solving, or agent tasks, and do not deeply analyze the failure sources in private-library code generation. In this setting, failures may arise not only from general task-solving mistakes, but also from incorrect use of individual APIs, improper coordination among multiple APIs, and flawed task-level solution strategies. Storing all experiences in a single memory space makes it difficult to distinguish these different types of reusable knowledge. Based on our failure analysis, \name introduces a multi-level evolving memory for private-library code generation, organizing execution feedback into API-level, cross-API-level, and task-level \textit{Usage Guidelines} to better support future code generation.

\section{Motivation}
\label{sec:motivation}


\begin{table}[t]
\centering
\small
\renewcommand{\arraystretch}{1.08}
\setlength{\tabcolsep}{5.2pt}
\caption{\textit{Pass@1} of Vanilla and Oracle settings on \ndonnx and \numba.}
\vspace{-0.1in}
\label{tab:oracle-study}
\resizebox{\columnwidth}{!}{
\begin{tabular}{l|ccc|ccc}
\toprule
\multirow{2}{*}{\textbf{Model}} &
\multicolumn{3}{c|}{\textbf{\ndonnx}} &
\multicolumn{3}{c}{\textbf{\numba}} \\
\cmidrule(lr){2-4}\cmidrule(lr){5-7}
& \textbf{Van.} & \textbf{Ora.} & \textbf{Gain}
& \textbf{Van.} & \textbf{Ora.} & \textbf{Gain} \\
\midrule
\llmname{Qwen2.5-7B}
& 12.19 & 44.79 & \cellcolor[HTML]{FFF4D6}\textbf{+32.60}
& 26.10 & 27.43 & \cellcolor[HTML]{FFF4D6}\textbf{+1.33} \\

\llmname{Qwen3-32B}
& 24.26 & 51.48 & \cellcolor[HTML]{FFF4D6}\textbf{+27.22}
& 52.94 & 53.48 & \cellcolor[HTML]{FFF4D6}\textbf{+0.54} \\

\llmname{GLM-4-Plus}
& 34.91 & 56.21 & \cellcolor[HTML]{FFF4D6}\textbf{+21.30}
& 52.41 & 56.68 & \cellcolor[HTML]{FFF4D6}\textbf{+4.27} \\
\bottomrule
\end{tabular}
}
\vspace{-0.15in}
\end{table}

\looseness=-1
To examine the upper bound of the RAG method under ideal retrieval conditions, we conduct a Vanilla--Oracle comparison. In the Vanilla setting, the model receives only the task requirement and starter code without additional API documentation. In the Oracle setting, all API documents required by the current task are directly injected, thereby eliminating retrieval errors and exposing the actual benefit of static API documentation. To avoid drawing conclusions from a single model capability level, we select three models with increasing capabilities and scales: Qwen2.5-Coder-7B-Instruct, Qwen3-32B, and GLM-4-Plus. For compactness, Table~\ref{tab:oracle-study} abbreviates them as Qwen2.5-7B, Qwen3-32B, and GLM-4-Plus, respectively. GLM-4-Plus does not disclose its parameter size, but can be regarded as a large proprietary model according to its official description. All experiments are conducted with temperature set to 0 to reduce sampling randomness.

\looseness=-1
As shown in Table~\ref{tab:oracle-study}, Oracle documentation indeed brings substantial gains on \ndonnx, improving \pass{1} by 32.60, 27.22, and 21.30 percentage points for the three models, respectively. However, even when retrieval errors are completely removed, the \pass{1} scores of all models remain below 60\%, suggesting that improving API recall alone is insufficient for reliable private-library code generation. On \numba, the gains from Oracle documentation are even more limited, with improvements of only 1.33, 0.54, and 4.27 percentage points. Further inspection shows that many \numba tasks already expose strong CUDA/Numba programming cues in their requirements and starter code, allowing larger models to solve part of the benchmark using general CUDA programming knowledge under the Vanilla setting. Nevertheless, adding Oracle API documentation still leads to only limited improvements. Overall, static API documentation helps LLMs identify \textit{what APIs to use}, but remains insufficient for teaching them \textit{how to use them correctly}.

\begin{boxK}
\small \faIcon{search} \textbf{Observation 1:}
Even with perfect API-document retrieval, static documentation remains insufficient for reliable private-library code generation, indicating that RAG method does not fully address the \textit{how-to-use} problem.
\end{boxK}


\looseness=-1
To understand why correct generation cannot be guaranteed even under Oracle settings, we perform a qualitative analysis of representative failure cases. As shown in Figure~\ref{fig:cases}, these failures suggest that LLMs still lack practical knowledge of how to use private-library APIs in specific tasks. We categorize them into three levels: API-level errors, cross-API-level errors, and task-level errors.

\looseness=-1
At the API level, even with relevant API documentation, LLMs may still incorrectly invoke individual APIs or generate unavailable APIs. Figure~\ref{fig:cases}(a) shows such an example. The model treats \texttt{ndonnx.Array} as a normal Python sequence and calls \texttt{len(x)} to check its length. However, \texttt{ndonnx.Array} does not support this operation, leading to the execution error \texttt{TypeError: object of type 'Array' has no len()}. The generated code also contains the unavailable API \texttt{ndx.average}, which is not included in the Oracle documentation. This case shows that static API documents cannot fully prevent incorrect object usage or API hallucination. 

At the Cross-API level, the problem lies not in a single API, but in how intermediate results flow across APIs. In Figure~\ref{fig:cases}(b), the model builds a data-flow chain from \texttt{equal} to \texttt{where} and then to \texttt{divide}, but the mask shape is not aligned with row-wise values, causing a malformed denominator to flow into \texttt{divide}.

At the Task level, LLMs may fail to organize the provided APIs into a solution that satisfies the task objective. Figure~\ref{fig:cases}(c) illustrates such a case: although the model uses APIs such as \texttt{sum}, \texttt{divide}, \texttt{multiply}, and \texttt{where}, it incorrectly treats \texttt{sum} as the row mean, making the subsequent scaling factor and final output inconsistent with the task requirement. These cases indicate that the remaining failures are not simply retrieval failures, but reusable \textit{how-to-use} errors at different levels.

\begin{boxK}
\small \faIcon{search} \textbf{Observation 2:}
Failures under Oracle documentation reveal reusable \textit{how-to-use} errors at three levels: individual API invocation, cross-API data flow, and task-solving strategy.
\end{boxK}

\section{Methodology}
\label{sec:method}


\begin{figure*}[t]
    \centering
    \setlength{\abovecaptionskip}{0pt}
    \setlength{\belowcaptionskip}{0pt}
    \includegraphics[
        width=\textwidth,
        trim={1cm 1.4cm 1cm 1cm},
        clip
    ]{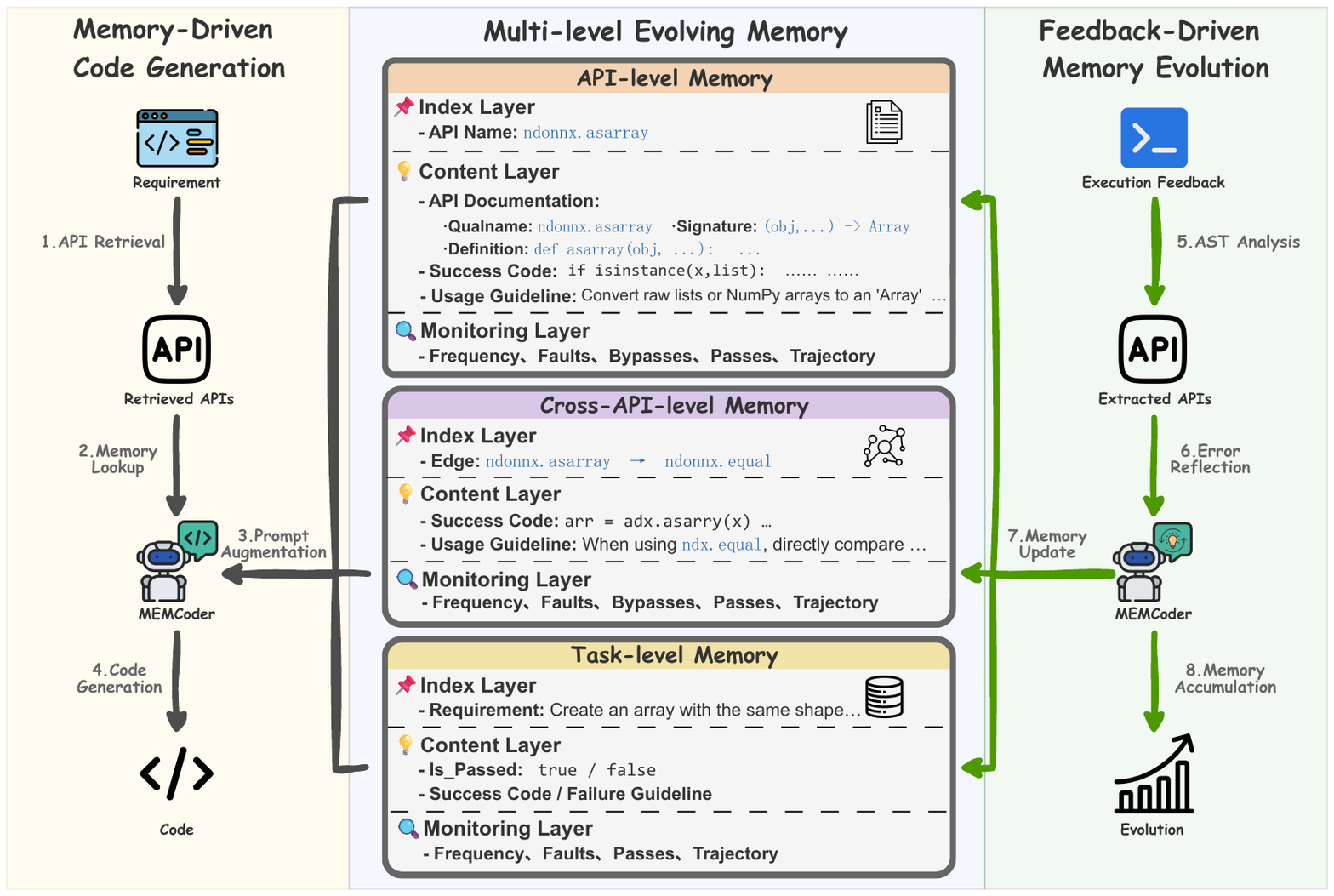}
\caption{Overview of the \textsc{MEMCoder} framework. \textbf{Middle:} The Multi-level Evolving Memory (\ref{sec:method-memory}) maintains API-level, Cross-API-level, and Task-level memories. \textbf{Left:} The Memory-Driven Code Generation pipeline (\ref{sec:method-generation}) retrieves API documents and relevant memories to augment the prompt for code generation. \textbf{Right:} The Feedback-Driven Memory Evolution module (\ref{sec:method-evolution}) uses execution feedback to update and accumulate reusable memories for future tasks.}
    \vspace{-14pt}
    \label{fig:overview}
\end{figure*}


\subsection{Overview}
\label{sec:method-overview}
\looseness=-1
We propose \name, a training-free self-evolving memory framework for private-library code generation. Given a target private library $\mathcal{L}$ and its static API documentation set $\mathcal{D}$, the goal is to generate a correct code solution $c_t$ for each requirement $r_t$ without updating the LLM parameters. The key idea is to augment existing RAG pipelines with an external self-evolving memory $\mathcal{M}$ that stores and reuses execution-derived memories from previous generations. As shown in Figure~\ref{fig:overview}, $\mathcal{M}$ consists of API-level memory $\mathcal{M}_{\text{API}}$, Cross-API-level memory $\mathcal{M}_{\text{Cross}}$, and Task-level memory $\mathcal{M}_{\text{Task}}$, which capture individual API usage, API coordination, and task-level solution experience, respectively.

\looseness=-1
The framework follows a closed-loop workflow with two phases. In the forward phase, Memory-Driven Code Generation (Section~\ref{sec:method-generation}), a base RAG method retrieves relevant API documents from $\mathcal{D}$. The retrieved APIs and the current requirement are then used to query $\mathcal{M}$, and both static documentation and retrieved memories are injected into the prompt to generate $c_t$. In the backward phase, Feedback-Driven Memory Evolution (Section~\ref{sec:method-evolution}), the generated code is executed to obtain feedback $f_t$. Based on the execution result, AST analysis, and error reflection, the corresponding task, API, and cross-API memories are updated. Through this generate--execute--feedback--update loop, the LLM gradually develops familiarity with the target private library without retraining.

\subsection{Multi-level Evolving Memory}
\label{sec:method-memory}
\looseness=-1
As shown in the middle part of Figure~\ref{fig:overview}, each memory entry contains three layers: an Index Layer for retrieval or matching, a Content Layer for prompt injection, and a Monitoring Layer for tracking later usage outcomes. To avoid unbounded growth, each API and each cross-API edge maintains only one successful code memory and one current usage guideline, while each historical task is stored as a compact entry.

\looseness=-1
\bi{API-Level Memory.}
API-level memory uses each documented private-library API name as its index. Its content includes the static API documentation, the most recent successful code involving this API, and the current usage guideline reflected from previous failures. The code provides an executable invocation example, while the guideline records how the API was misused and what should be avoided or corrected in later generations. Invalid API name warnings are also maintained for hallucinated API names that appear in generated code but are absent from the documentation. Each API memory has a monitoring record with $Freq$, $Pass$, $Fault$, $Bypass$, and $Traj$, recording how often the memory is used, how often the task succeeds after its use, how often a failed task is judged to contain an error in this API usage, how often a task fails while this API usage is not judged erroneous, and the chronological sequence of usage outcomes.

\looseness=-1
\bi{Cross-API-Level Memory.}
Cross-API-level memory is organized as a directed API coordination graph. Each edge is indexed by an ordered API pair $(a_i, a_j)$, representing an observed data-flow or coordination relation from $a_i$ to $a_j$. Its content stores the latest successful code slice demonstrating this relation and the current usage guideline for this API pair. Cross-API memory uses the same monitoring fields as API-level memory.

\looseness=-1
\bi{Task-Level Memory.}
Task-level memory stores historical tasks as compact entries indexed by task requirements. A successful task stores its passing code, while a failed task stores the failure guideline reflected from execution feedback. Its monitoring record contains $Freq$, $Pass$, $Fault$, and $Traj$, where $Pass$ and $Fault$ count the successful and failed sampled generations observed after this memory is reused. Since bypass decisions are less clear for task memories than for API-level or Cross-API-level memories, Task-level memory does not distinguish bypass cases.

\subsection{Memory-Driven Code Generation}
\label{sec:method-generation}

Given the current requirement $r_t$, a base RAG method first retrieves candidate API documents $\mathcal{D}_{r_t}$ from $\mathcal{D}$ and obtains the corresponding API set $\mathcal{A}_{r_t}$. The final prompt is constructed by combining the retrieved documents with three types of memories:
\begin{equation}
P_t = \textsc{Prompt}(\mathcal{D}_{r_t},
\mathcal{M}^{\text{API}}_{r_t},
\mathcal{M}^{\text{Cross}}_{r_t},
\mathcal{M}^{\text{Task}}_{r_t}, r_t).
\end{equation}
where $\mathcal{M}^{\text{API}}_{r_t}$, $\mathcal{M}^{\text{Cross}}_{r_t}$, and $\mathcal{M}^{\text{Task}}_{r_t}$ denote the three types of memories retrieved for the current task. The LLM then generates the final code $c_t$ from this augmented prompt.

\looseness=-1
API-level memory is matched directly using $\mathcal{A}_{r_t}$. For each retrieved API, its successful code and usage guideline are appended to the corresponding documentation block. If a hallucinated API name has exceeded a predefined failure threshold or previously co-occurred with any current candidate API, an additional invalid API name warning is injected.

\looseness=-1
Cross-API-level memory is retrieved from internal edges among the candidate APIs. If an edge $(a_i, a_j)$ exists in the coordination graph and both endpoints belong to $\mathcal{A}_{r_t}$, its successful code slice and usage guideline are injected into the prompt.

\looseness=-1
Task-level memory is retrieved by semantic similarity between the current requirement $r_t$ and historical requirements $r_i$, weighted by the historical utility of each memory:
\begin{equation}
\operatorname{score}(m_i, r_t)
=
sim(r_t, r_i)
\cdot
\frac{Pass_i + 1}{Pass_i + Fault_i + 2}.
\end{equation}
where $sim(\cdot,\cdot)$ denotes embedding-based cosine similarity. The additive constants serve as a smoothing prior: a newly stored task memory starts with a neutral utility before sufficient reuse feedback is accumulated, rather than being treated as fully reliable. As the memory is reused, successful sampled generations increase its utility, while failed sampled generations gradually reduce its ranking score. The top-ranked task memories are injected into the prompt: successful task memories provide requirements and passing code, while failed task memories provide requirements and failure guidelines.

\subsection{Feedback-Driven Memory Evolution}
\label{sec:method-evolution}

\looseness=-1
After generation, the code $c_t$ is executed to obtain feedback $f_t$, including test outcomes, runtime errors, assertion failures, or other execution information. This phase converts the current generation and feedback into new memories, while updating the historical memories used in the current round.

\looseness=-1
\bi{API-Level Evolution.}
AST analysis first extracts the private-library APIs actually used in $c_t$. If the task succeeds, the successful code related to each documented API is stored or updated. If the task fails, the failure is not attributed to all used APIs by default. Instead, for each documented API, \name invokes an LLM-based error attributor to judge whether this API is misused, based on the requirement, generated code, execution feedback, API documentation, and the existing guideline. If the API is judged as the error source, an LLM-based reflector further summarizes the failure and revises the API-level usage guideline, while $Fault$ is increased. Otherwise, the API memory is marked as bypass. Private-library API names that appear in the code but are absent from the documentation are recorded as invalid API name warnings, together with their co-occurring documented APIs.

\looseness=-1
\bi{Cross-API-Level Evolution.}
AST analysis also extracts data-flow relations among APIs. New cross-API edges are created only from successful executions, since API combinations in failed code may not represent valid coordination patterns. For existing edges, successful tasks update the code slice and monitoring record. When the task fails, \name invokes the LLM-based error attributor only for existing edges to determine whether the failure is caused by incorrect API coordination. If an edge is judged as faulty, the LLM-based reflector revises its cross-API usage guideline and $Fault$ is increased; otherwise, the edge is marked as bypass.

\looseness=-1
\bi{Task-Level Evolution.}
For each completed task, a compact task memory is written. Successful tasks store the requirement and passing code, while failed tasks store the requirement and the reflected failure guideline. For task memories used in the current generation, the monitoring record is updated according to the final task outcome by increasing $Pass$ or $Fault$ and appending the result to $Traj$.

\section{Experimental Setup}
\label{sec:setup}

\looseness=-1
To assess \name, we conduct comprehensive experiments to answer three Research Questions (RQs). In this section, we present the details of our experimental setup.

\subsection{Research Questions}
\label{sec:rq}

\looseness=-1
\textbf{RQ1: Can \name effectively enhance RAG-based private-library code generation?}
This RQ evaluates whether \name can serve as an effective enhancement to existing RAG pipelines. To answer it, we apply \name on top of different API retrieval methods, including Naive RAG, EpiGEN, and CAPIR, and examine whether self-evolving memory accumulated from execution feedback can further improve code generation beyond static API documentation.

\looseness=-1
\textbf{RQ2: How does \name compare with existing self-evolving memory methods?}
This RQ evaluates whether the self-evolving memory framework of \name is more effective than existing general memory-based methods for private-library code generation. To answer it, we compare \name with representative methods that also accumulate and reuse past experience during inference, and examine whether the closed-loop process of memory-driven generation and feedback-driven evolution leads to better adaptation over a continuous task stream.


\looseness=-1
\textbf{RQ3: What are the individual contributions of the three memory levels in \name?}
This RQ analyzes the contribution of each memory component to overall performance. To answer this question, we remove API-level memory, Cross-API-level memory, and Task-level memory individually and compare these variants with the full version of \name.

\subsection{Benchmarks}
\label{sec:benchmark}

\looseness=-1
We evaluate \name on two private-library code generation benchmarks, \textbf{NdonnxEval} and \textbf{NumbaEval}~\cite{zhang2026masterteachingllmsuse}. Since real proprietary libraries are usually inaccessible, these benchmarks use recently released public libraries as proxies. Both target libraries were released in 2024 and actively developed in 2025, reducing the chance that models have memorized their latest APIs or benchmark tasks.

\begin{itemize}[leftmargin=15pt]
    \item \textbf{NdonnxEval} evaluates code generation with \texttt{ndonnx}, an ONNX-based tensor library. It contains 169 manually curated tasks, each requiring more than 4 APIs on average and verified by over 9 unit tests.
    
    \item \textbf{NumbaEval} targets \texttt{numba-cuda} for CUDA JIT programming. It contains 187 manually curated tasks with complex algorithmic requirements, also requiring more than 4 APIs on average and verified by over 9 unit tests.
\end{itemize}

\subsection{Metrics}
\label{sec:metric}

\looseness=-1
We use functional correctness and efficiency as our main evaluation metrics. To reduce randomness and obtain more reliable estimates, we compute the functional metrics using standard unbiased estimators.

\begin{itemize}[leftmargin=12pt]
    \item \textbf{Pass@k ($k \in \{1, 3, 5\}$).} For each instance, we sample $n \ge k$ candidate solutions (we use $n=10$), execute the provided test cases, and count the number of passing solutions $c$. Following prior work~\cite{chen2022codet, athiwaratkun2022multi}, we compute Pass@k using the unbiased estimator:
    \begin{equation}
    \text{Pass@k}=\mathbb{E}_{\text{instances}}\!\left[1-\frac{\binom{n-c}{k}}{\binom{n}{k}}\right].
    \end{equation}

    \item \textbf{Exec@k ($k \in \{1, 3, 5\}$).} As models often misuse private APIs and trigger runtime failures, we report Exec@k to measure basic executability. It is defined analogously to Pass@k, except that a solution is counted as successful if it runs to completion on the test inputs without raising any runtime exceptions.

\end{itemize}

\subsection{Baselines}
\label{sec:baseline}
\looseness=-1
We compare \name with two categories of baselines: RAG-based methods designed for private-library code generation, and representative self-evolving memory methods that accumulate and reuse historical experience during inference.

\looseness=-1
\textit{(1) RAG-based Methods.}
To evaluate whether \name can enhance existing API retrieval pipelines, we select three representative RAG-based methods as backbones and compare their performance before and after integrating \name.
\textbf{Naive RAG}~\cite{docprompting, apifinder} follows the standard RAG paradigm: it embeds API documentation, retrieves relevant APIs according to the task requirement, and injects their static documentation into the prompt.
\textbf{EpiGEN}~\cite{epigen} improves API retrieval by decomposing complex requirements into fine-grained subtasks and retrieving APIs for each subtask.
\textbf{CAPIR}~\cite{capir} further improves retrieval by decomposing requirements and reranking candidate APIs to obtain more accurate API contexts.
By applying \name on top of these backbones, we examine whether multi-level evolving memory can consistently improve different RAG pipelines.

\looseness=-1
\textit{(2) Self-evolving Memory Methods.}
To evaluate whether the memory design of \name is more suitable for private-library code generation, we compare it with representative self-evolving memory methods that also accumulate and reuse experience across a task stream. Since these methods are not originally designed for private-library code generation, we adapt them by equipping each method with the same Naive RAG module for retrieving and injecting relevant API documentation.
\textbf{Dynamic Cheatsheet (DC-RS)}~\cite{suzgun2025dynamic} maintains an adaptive external memory of reusable problem-solving strategies and code snippets, and uses an LLM curator to generate a task-specific cheatsheet from relevant historical experiences before generation.
\textbf{ReMem}~\cite{wei2025evomem} stores historical requirements, generated outputs, and execution feedback as unified memory entries. For each new task, it retrieves similar past experiences and follows a ``Think-Act-Refine'' loop, where the LLM reasons about the current requirement, inspects the retrieved memories, and refines its solution based on historical feedback.
\begin{table*}[!t]
\centering
\small
\renewcommand{\arraystretch}{1.1}
\setlength{\tabcolsep}{4pt}
\caption{Performance comparison with \textbf{RAG-based baselines} in terms of \pass{k} and \exec{k} (\%) on \ndonnx and \numba.}
\vspace{-0.1in}
\label{tab:rag_comparison}
\scalebox{0.9}{
\begin{tabular}{c|c|cccccc|cccccc}
\toprule
\multirow[c]{2}{*}[-0.45ex]{\textbf{Model}}
& \multirow[c]{2}{*}[-0.45ex]{\textbf{Method}}
& \multicolumn{6}{c|}{\textbf{\ndonnx}}
& \multicolumn{6}{c}{\textbf{\numba}} \\
\cmidrule(lr){3-8}\cmidrule(lr){9-14}
& & \textbf{\pass{1}} & \textbf{\pass{3}} & \textbf{\pass{5}} & \textbf{\exec{1}} & \textbf{\exec{3}} & \textbf{\exec{5}}
  & \textbf{\pass{1}} & \textbf{\pass{3}} & \textbf{\pass{5}} & \textbf{\exec{1}} & \textbf{\exec{3}} & \textbf{\exec{5}} \\
\midrule

\multirow{6}{*}{\shortstack{\fontfamily{pcr}\selectfont Qwen2.5\\\fontfamily{pcr}\selectfont-Coder}}
& \textit{Naive RAG}
& 25.21 & 37.49 & 42.62 & 33.25 & 47.70 & 52.86
& 23.16 & 40.04 & 48.09 & 40.27 & 66.37 & 76.82 \\
& \cellcolor[HTML]{FFF4D6}+\name
& \cellcolor[HTML]{FFF4D6}\textbf{41.78} & \cellcolor[HTML]{FFF4D6}\textbf{52.72} & \cellcolor[HTML]{FFF4D6}\textbf{56.87} & \cellcolor[HTML]{FFF4D6}\textbf{50.36} & \cellcolor[HTML]{FFF4D6}\textbf{62.93} & \cellcolor[HTML]{FFF4D6}\textbf{67.59}
& \cellcolor[HTML]{FFF4D6}\textbf{32.46} & \cellcolor[HTML]{FFF4D6}\textbf{45.41} & \cellcolor[HTML]{FFF4D6}\textbf{50.54} & \cellcolor[HTML]{FFF4D6}\textbf{51.28} & \cellcolor[HTML]{FFF4D6}\textbf{70.76} & \cellcolor[HTML]{FFF4D6}\textbf{77.03} \\
& \textit{EpiGen}
& 23.61 & 36.49 & 41.74 & 30.24 & 46.05 & 52.39
& 18.98 & 35.49 & 44.20 & 33.85 & 60.42 & 71.64 \\
& \cellcolor[HTML]{FFF4D6}+\name
& \cellcolor[HTML]{FFF4D6}\textbf{44.91} & \cellcolor[HTML]{FFF4D6}\textbf{56.43} & \cellcolor[HTML]{FFF4D6}\textbf{60.74} & \cellcolor[HTML]{FFF4D6}\textbf{54.02} & \cellcolor[HTML]{FFF4D6}\textbf{66.62} & \cellcolor[HTML]{FFF4D6}\textbf{71.16}
& \cellcolor[HTML]{FFF4D6}\textbf{30.37} & \cellcolor[HTML]{FFF4D6}\textbf{43.79} & \cellcolor[HTML]{FFF4D6}\textbf{49.46} & \cellcolor[HTML]{FFF4D6}\textbf{49.89} & \cellcolor[HTML]{FFF4D6}\textbf{70.68} & \cellcolor[HTML]{FFF4D6}\textbf{77.64} \\
& \textit{CAPIR}
& 31.54 & 45.98 & 51.08 & 39.47 & 56.04 & 61.42
& 21.02 & 38.12 & 45.99 & 35.45 & 61.67 & 72.18 \\
& \cellcolor[HTML]{FFF4D6}+\name
& \cellcolor[HTML]{FFF4D6}\textbf{51.42} & \cellcolor[HTML]{FFF4D6}\textbf{62.09} & \cellcolor[HTML]{FFF4D6}\textbf{66.43} & \cellcolor[HTML]{FFF4D6}\textbf{61.12} & \cellcolor[HTML]{FFF4D6}\textbf{71.75} & \cellcolor[HTML]{FFF4D6}\textbf{75.40}
& \cellcolor[HTML]{FFF4D6}\textbf{34.97} & \cellcolor[HTML]{FFF4D6}\textbf{48.03} & \cellcolor[HTML]{FFF4D6}\textbf{53.93} & \cellcolor[HTML]{FFF4D6}\textbf{57.06} & \cellcolor[HTML]{FFF4D6}\textbf{74.02} & \cellcolor[HTML]{FFF4D6}\textbf{78.61} \\
          
\midrule

\multirow{6}{*}{\shortstack{\fontfamily{pcr}\selectfont Llama-3.1}}
& \textit{Naive RAG}
& 11.48 & 19.84 & 24.64 & 24.26 & 41.01 & 49.25
& 6.31 & 14.27 & 19.49 & 23.58 & 49.31 & 62.90 \\
& \cellcolor[HTML]{FFF4D6}+\name
& \cellcolor[HTML]{FFF4D6}\textbf{34.50} & \cellcolor[HTML]{FFF4D6}\textbf{47.22} & \cellcolor[HTML]{FFF4D6}\textbf{52.67} & \cellcolor[HTML]{FFF4D6}\textbf{47.63} & \cellcolor[HTML]{FFF4D6}\textbf{64.39} & \cellcolor[HTML]{FFF4D6}\textbf{71.21}
& \cellcolor[HTML]{FFF4D6}\textbf{45.99} & \cellcolor[HTML]{FFF4D6}\textbf{61.71} & \cellcolor[HTML]{FFF4D6}\textbf{68.27} & \cellcolor[HTML]{FFF4D6}\textbf{77.97} & \cellcolor[HTML]{FFF4D6}\textbf{91.39} & \cellcolor[HTML]{FFF4D6}\textbf{94.39} \\
& \textit{EpiGen}
& 11.72 & 21.51 & 26.02 & 20.95 & 36.24 & 43.33
& 6.90 & 16.60 & 22.95 & 24.06 & 49.56 & 62.70 \\
& \cellcolor[HTML]{FFF4D6}+\name
& \cellcolor[HTML]{FFF4D6}\textbf{22.66} & \cellcolor[HTML]{FFF4D6}\textbf{33.42} & \cellcolor[HTML]{FFF4D6}\textbf{38.22} & \cellcolor[HTML]{FFF4D6}\textbf{35.50} & \cellcolor[HTML]{FFF4D6}\textbf{51.34} & \cellcolor[HTML]{FFF4D6}\textbf{57.56}
& \cellcolor[HTML]{FFF4D6}\textbf{18.34} & \cellcolor[HTML]{FFF4D6}\textbf{30.52} & \cellcolor[HTML]{FFF4D6}\textbf{35.98} & \cellcolor[HTML]{FFF4D6}\textbf{46.15} & \cellcolor[HTML]{FFF4D6}\textbf{69.34} & \cellcolor[HTML]{FFF4D6}\textbf{78.28} \\
& \textit{CAPIR}
& 13.31 & 22.90 & 27.83 & 21.95 & 38.63 & 46.75
& 5.13 & 12.84 & 18.30 & 19.89 & 43.89 & 57.61 \\
& \cellcolor[HTML]{FFF4D6}+\name
& \cellcolor[HTML]{FFF4D6}\textbf{26.75} & \cellcolor[HTML]{FFF4D6}\textbf{41.48} & \cellcolor[HTML]{FFF4D6}\textbf{48.11} & \cellcolor[HTML]{FFF4D6}\textbf{42.54} & \cellcolor[HTML]{FFF4D6}\textbf{60.36} & \cellcolor[HTML]{FFF4D6}\textbf{68.43}
& \cellcolor[HTML]{FFF4D6}\textbf{20.80} & \cellcolor[HTML]{FFF4D6}\textbf{33.38} & \cellcolor[HTML]{FFF4D6}\textbf{39.52} & \cellcolor[HTML]{FFF4D6}\textbf{44.97} & \cellcolor[HTML]{FFF4D6}\textbf{68.30} & \cellcolor[HTML]{FFF4D6}\textbf{77.82} \\
\midrule

\multirow{6}{*}{\shortstack{\fontfamily{pcr}\selectfont DeepSeek\\\fontfamily{pcr}\selectfont-Coder}}
& \textit{Naive RAG}
& 25.50 & 40.51 & 46.69 & 35.86 & 53.97 & 61.02
& 11.02 & 25.39 & 34.35 & 23.90 & 49.91 & 63.36 \\
& \cellcolor[HTML]{FFF4D6}+\name
& \cellcolor[HTML]{FFF4D6}\textbf{44.79} & \cellcolor[HTML]{FFF4D6}\textbf{58.65} & \cellcolor[HTML]{FFF4D6}\textbf{63.46} & \cellcolor[HTML]{FFF4D6}\textbf{56.21} & \cellcolor[HTML]{FFF4D6}\textbf{70.65} & \cellcolor[HTML]{FFF4D6}\textbf{75.10}
& \cellcolor[HTML]{FFF4D6}\textbf{27.86} & \cellcolor[HTML]{FFF4D6}\textbf{42.45} & \cellcolor[HTML]{FFF4D6}\textbf{48.14} & \cellcolor[HTML]{FFF4D6}\textbf{49.57} & \cellcolor[HTML]{FFF4D6}\textbf{74.67} & \cellcolor[HTML]{FFF4D6}\textbf{83.29} \\
& \textit{EpiGen}
& 30.06 & 44.75 & 50.15 & 38.40 & 56.40 & 63.06
& 11.07 & 24.12 & 32.09 & 24.33 & 50.62 & 64.77 \\
& \cellcolor[HTML]{FFF4D6}+\name
& \cellcolor[HTML]{FFF4D6}\textbf{47.51} & \cellcolor[HTML]{FFF4D6}\textbf{60.62} & \cellcolor[HTML]{FFF4D6}\textbf{64.88} & \cellcolor[HTML]{FFF4D6}\textbf{58.22} & \cellcolor[HTML]{FFF4D6}\textbf{72.88} & \cellcolor[HTML]{FFF4D6}\textbf{76.24}
& \cellcolor[HTML]{FFF4D6}\textbf{32.94} & \cellcolor[HTML]{FFF4D6}\textbf{48.20} & \cellcolor[HTML]{FFF4D6}\textbf{54.53} & \cellcolor[HTML]{FFF4D6}\textbf{52.25} & \cellcolor[HTML]{FFF4D6}\textbf{75.57} & \cellcolor[HTML]{FFF4D6}\textbf{83.15} \\
& \textit{CAPIR}
& 25.92 & 41.92 & 48.34 & 36.39 & 55.44 & 62.81
& 10.11 & 23.00 & 31.48 & 21.60 & 45.74 & 59.06 \\
& \cellcolor[HTML]{FFF4D6}+\name
& \cellcolor[HTML]{FFF4D6}\textbf{42.25} & \cellcolor[HTML]{FFF4D6}\textbf{56.50} & \cellcolor[HTML]{FFF4D6}\textbf{61.29} & \cellcolor[HTML]{FFF4D6}\textbf{55.21} & \cellcolor[HTML]{FFF4D6}\textbf{70.66} & \cellcolor[HTML]{FFF4D6}\textbf{75.08}
& \cellcolor[HTML]{FFF4D6}\textbf{26.47} & \cellcolor[HTML]{FFF4D6}\textbf{37.14} & \cellcolor[HTML]{FFF4D6}\textbf{41.11} & \cellcolor[HTML]{FFF4D6}\textbf{46.58} & \cellcolor[HTML]{FFF4D6}\textbf{67.82} & \cellcolor[HTML]{FFF4D6}\textbf{75.37} \\
\bottomrule
\end{tabular}
}
\vspace{-0.15in}
\end{table*}
\begin{table}[t]
\centering
\small
\renewcommand{\arraystretch}{1.1}
\setlength{\tabcolsep}{4.5pt}
\caption{Performance comparison with \textbf{RAG-based baselines} in terms of \pass{1} and \exec{1} (\%) on \ndonnx and \numba.}
\vspace{-0.1in}
\label{tab:api_rag_comparison}
\resizebox{\columnwidth}{!}{
\begin{tabular}{c|c|cc|cc}
\toprule
\multirow[c]{2}{*}[-0.45ex]{\textbf{Model}}
& \multirow[c]{2}{*}[-0.45ex]{\textbf{Method}}
& \multicolumn{2}{c|}{\textbf{\ndonnx}}
& \multicolumn{2}{c}{\textbf{\numba}} \\
\cmidrule(lr){3-4}\cmidrule(lr){5-6}
& & \textbf{\pass{1}} & \textbf{\exec{1}}
  & \textbf{\pass{1}} & \textbf{\exec{1}} \\
\midrule

\multirow[c]{6}{*}[-0.35ex]{\shortstack{\fontfamily{pcr}\selectfont Qwen3\\\fontfamily{pcr}\selectfont-32B}}
& \textit{Naive RAG}
& 31.36 & 40.24
& 49.73 & 68.45 \\
& \cellcolor[HTML]{FFF4D6}+\name
& \cellcolor[HTML]{FFF4D6}\textbf{49.70}
& \cellcolor[HTML]{FFF4D6}\textbf{57.40}
& \cellcolor[HTML]{FFF4D6}\textbf{68.45}
& \cellcolor[HTML]{FFF4D6}\textbf{87.17} \\
& \textit{EpiGEN}
& 18.93 & 28.40
& 50.80 & 63.64 \\
& \cellcolor[HTML]{FFF4D6}+\name
& \cellcolor[HTML]{FFF4D6}\textbf{49.11}
& \cellcolor[HTML]{FFF4D6}\textbf{53.25}
& \cellcolor[HTML]{FFF4D6}\textbf{63.64}
& \cellcolor[HTML]{FFF4D6}\textbf{81.82} \\
& \textit{CAPIR}
& 32.54 & 38.46
& 54.55 & 65.24 \\
& \cellcolor[HTML]{FFF4D6}+\name
& \cellcolor[HTML]{FFF4D6}\textbf{62.72}
& \cellcolor[HTML]{FFF4D6}\textbf{71.60}
& \cellcolor[HTML]{FFF4D6}\textbf{65.78}
& \cellcolor[HTML]{FFF4D6}\textbf{81.82} \\

\midrule

\multirow[c]{6}{*}[-0.35ex]{\shortstack{\fontfamily{pcr}\selectfont GLM-4\\\fontfamily{pcr}\selectfont-Plus}}
& \textit{Naive RAG}
& 26.63 & 39.05
& 54.55 & 73.26 \\
& \cellcolor[HTML]{FFF4D6}+\name
& \cellcolor[HTML]{FFF4D6}\textbf{52.66}
& \cellcolor[HTML]{FFF4D6}\textbf{60.36}
& \cellcolor[HTML]{FFF4D6}\textbf{71.12}
& \cellcolor[HTML]{FFF4D6}\textbf{92.51} \\
& \textit{EpiGEN}
& 34.91 & 46.75
& 49.20 & 66.84 \\
& \cellcolor[HTML]{FFF4D6}+\name
& \cellcolor[HTML]{FFF4D6}\textbf{57.40}
& \cellcolor[HTML]{FFF4D6}\textbf{67.46}
& \cellcolor[HTML]{FFF4D6}\textbf{51.34}
& \cellcolor[HTML]{FFF4D6}\textbf{67.91} \\
& \textit{CAPIR}
& 33.14 & 45.56
& 54.01 & 73.26 \\
& \cellcolor[HTML]{FFF4D6}+\name
& \cellcolor[HTML]{FFF4D6}\textbf{69.23}
& \cellcolor[HTML]{FFF4D6}\textbf{73.37}
& \cellcolor[HTML]{FFF4D6}\textbf{66.84}
& \cellcolor[HTML]{FFF4D6}\textbf{84.49} \\

\bottomrule
\end{tabular}
}
\vspace{-0.15in}
\end{table}

\subsection{Models}
\label{sec:model}

\looseness=-1
Private-library code generation often involves confidential project code and internal APIs, so enterprise deployment commonly favors locally hosted models. Accordingly, we select three widely used open-source code LLMs at the 7B--8B scale that are practical for local deployment and have limited prior exposure to the target libraries: \textbf{Qwen2.5-Coder-7B-Instruct}~\cite{hui2024qwen25coder}, \textbf{Llama-3.1-8B-Instruct}~\cite{llama}, and \textbf{DeepSeek-Coder-6.7B-Instruct}~\cite{deepseekcoder}. For brevity, we refer to them as \texttt{Qwen2.5-Coder}, \texttt{Llama-3.1}, and \texttt{DeepSeek-Coder}. To avoid limiting our conclusions to small-scale locally deployed models, we further evaluate two larger LLMs through API calls: \textbf{Qwen3-32B}~\cite{yang2025qwen3} and \textbf{GLM-4-Plus}~\cite{glm2024chatglm}. These additional models allow us to examine whether the observed benefits of self-evolving memory persist when the backbone LLM has stronger general coding capability.

\begin{figure}[t]
    \centering
    \setlength{\abovecaptionskip}{2pt}
    \setlength{\belowcaptionskip}{0pt}
    \includegraphics[
        width=\columnwidth,
        trim={1cm 5.5cm 1.2cm 1cm},
        clip
    ]{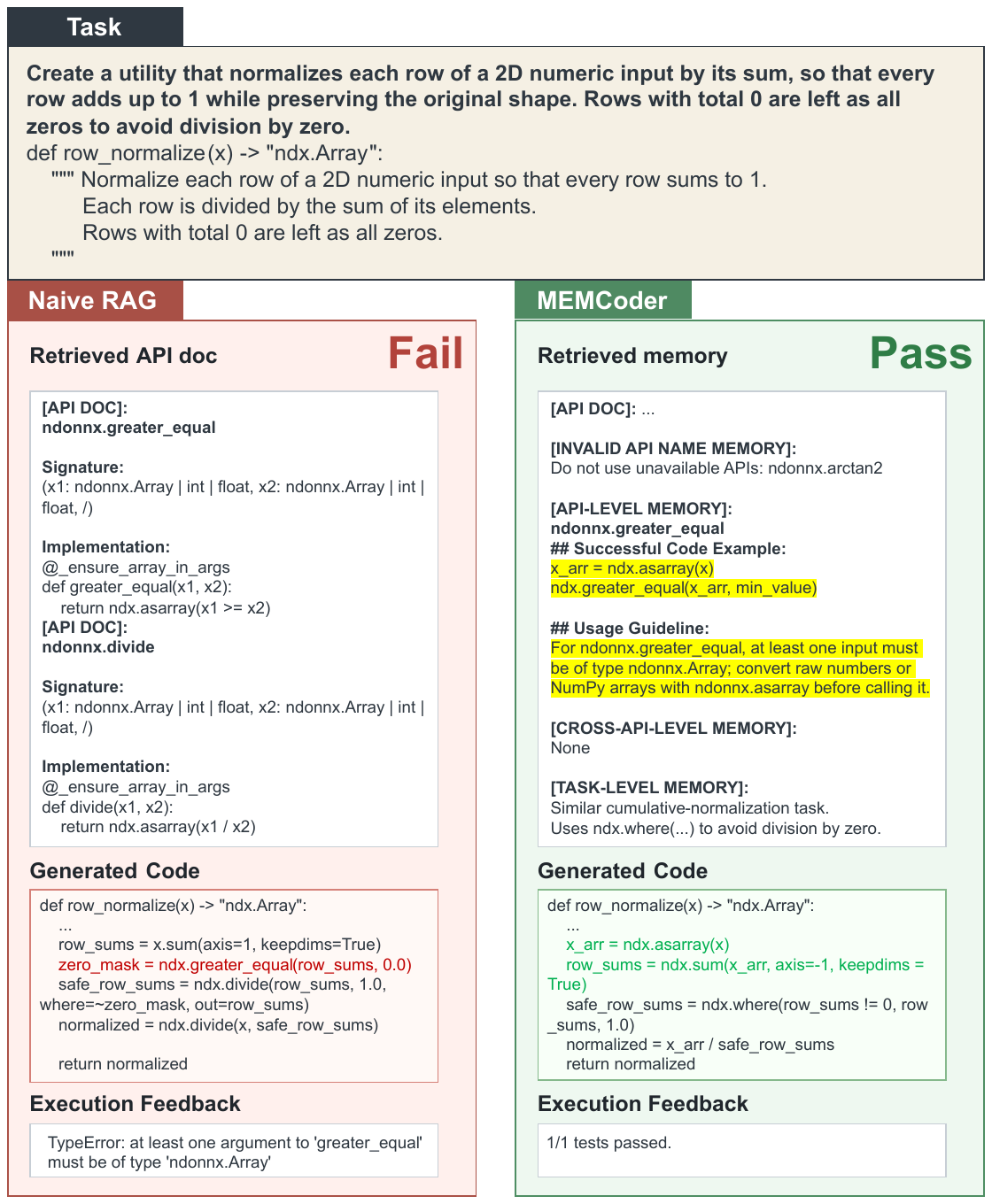}
    \caption{Case study of memory-guided API usage in \name.}
    \label{fig:rq3_memory_reuse_case}
    \vspace{-0.1in}
\end{figure}

\begin{table*}[!t]
\centering
\small
\renewcommand{\arraystretch}{1.1}
\setlength{\tabcolsep}{4pt}
\caption{Performance comparison with \textbf{self-evolving memory baselines} in terms of \pass{k} and \exec{k} (\%) on \ndonnx and \numba.}
\vspace{-0.1in}
\label{tab:cl_comparison}
\scalebox{0.9}{
\begin{tabular}{c|c|cccccc|cccccc}
\toprule
\multirow[c]{2}{*}[-0.45ex]{\textbf{Model}}
& \multirow[c]{2}{*}[-0.45ex]{\textbf{Method}}
& \multicolumn{6}{c|}{\textbf{\ndonnx}}
& \multicolumn{6}{c}{\textbf{\numba}} \\
\cmidrule(lr){3-8}\cmidrule(lr){9-14}
& & \textbf{\pass{1}} & \textbf{\pass{3}} & \textbf{\pass{5}} & \textbf{\exec{1}} & \textbf{\exec{3}} & \textbf{\exec{5}}
  & \textbf{\pass{1}} & \textbf{\pass{3}} & \textbf{\pass{5}} & \textbf{\exec{1}} & \textbf{\exec{3}} & \textbf{\exec{5}} \\
\midrule

\multirow[c]{3}{*}[-0.35ex]{\shortstack{\fontfamily{pcr}\selectfont Qwen2.5\\\fontfamily{pcr}\selectfont-Coder}}
& ReMem
& 21.42 & 34.02 & 39.73 & 28.70 & 44.84 & 51.68
& 22.78 & 39.70 & 48.48 & 40.11 & 66.44 & 76.35 \\
& DC-RS
& 22.01 & 41.35 & 50.51 & 27.75 & 49.88 & 59.71
& 13.58 & 26.63 & 34.60 & 26.36 & 50.01 & 61.40 \\
& \cellcolor[HTML]{FFF4D6}\name
& \cellcolor[HTML]{FFF4D6}\textbf{41.78}
& \cellcolor[HTML]{FFF4D6}\textbf{52.72}
& \cellcolor[HTML]{FFF4D6}\textbf{56.87}
& \cellcolor[HTML]{FFF4D6}\textbf{50.36}
& \cellcolor[HTML]{FFF4D6}\textbf{62.93}
& \cellcolor[HTML]{FFF4D6}\textbf{67.59}
& \cellcolor[HTML]{FFF4D6}\textbf{32.46}
& \cellcolor[HTML]{FFF4D6}\textbf{45.41}
& \cellcolor[HTML]{FFF4D6}\textbf{50.54}
& \cellcolor[HTML]{FFF4D6}\textbf{51.28}
& \cellcolor[HTML]{FFF4D6}\textbf{70.76}
& \cellcolor[HTML]{FFF4D6}\textbf{77.03} \\
\midrule

\multirow[c]{3}{*}[-0.35ex]{\shortstack{\fontfamily{pcr}\selectfont Llama-3.1}}
& ReMem
& 11.01 & 21.79 & 27.46 & 22.78 & 43.82 & 53.61
& 3.74 & 9.34 & 13.36 & 22.51 & 49.30 & 64.19 \\
& DC-RS
& 10.00 & 22.24 & 29.42 & 18.70 & 38.86 & 49.35
& 8.24 & 17.29 & 22.25 & 24.22 & 49.00 & 60.95 \\
& \cellcolor[HTML]{FFF4D6}\name
& \cellcolor[HTML]{FFF4D6}\textbf{34.50}
& \cellcolor[HTML]{FFF4D6}\textbf{47.22}
& \cellcolor[HTML]{FFF4D6}\textbf{52.67}
& \cellcolor[HTML]{FFF4D6}\textbf{47.63}
& \cellcolor[HTML]{FFF4D6}\textbf{64.39}
& \cellcolor[HTML]{FFF4D6}\textbf{71.21}
& \cellcolor[HTML]{FFF4D6}\textbf{45.99}
& \cellcolor[HTML]{FFF4D6}\textbf{61.71}
& \cellcolor[HTML]{FFF4D6}\textbf{68.27}
& \cellcolor[HTML]{FFF4D6}\textbf{77.97}
& \cellcolor[HTML]{FFF4D6}\textbf{91.39}
& \cellcolor[HTML]{FFF4D6}\textbf{94.39} \\
\midrule

\multirow[c]{3}{*}[-0.35ex]{\shortstack{\fontfamily{pcr}\selectfont DeepSeek\\\fontfamily{pcr}\selectfont-Coder}}
& ReMem
& 22.49 & 37.18 & 44.18 & 32.90 & 52.48 & 60.58
& 6.42 & 16.37 & 23.59 & 15.40 & 36.53 & 49.83 \\
& DC-RS
& 15.74 & 30.97 & 38.28 & 29.35 & 52.87 & 62.49
& 7.11 & 17.98 & 25.82 & 14.65 & 35.32 & 48.76 \\
& \cellcolor[HTML]{FFF4D6}\name
& \cellcolor[HTML]{FFF4D6}\textbf{44.79}
& \cellcolor[HTML]{FFF4D6}\textbf{58.65}
& \cellcolor[HTML]{FFF4D6}\textbf{63.46}
& \cellcolor[HTML]{FFF4D6}\textbf{56.21}
& \cellcolor[HTML]{FFF4D6}\textbf{70.65}
& \cellcolor[HTML]{FFF4D6}\textbf{75.10}
& \cellcolor[HTML]{FFF4D6}\textbf{27.86}
& \cellcolor[HTML]{FFF4D6}\textbf{42.45}
& \cellcolor[HTML]{FFF4D6}\textbf{48.14}
& \cellcolor[HTML]{FFF4D6}\textbf{49.57}
& \cellcolor[HTML]{FFF4D6}\textbf{74.67}
& \cellcolor[HTML]{FFF4D6}\textbf{83.29} \\
\bottomrule
\end{tabular}
}
\vspace{-0.15in}
\end{table*}
\begin{table}[t]
\centering
\small
\renewcommand{\arraystretch}{1.1}
\setlength{\tabcolsep}{4.5pt}
\caption{Performance comparison with \textbf{self-evolving memory baselines} in terms of \pass{1} and \exec{1} (\%) on \ndonnx and \numba.}
\vspace{-0.1in}
\label{tab:api_cl_comparison}
\resizebox{\columnwidth}{!}{
\begin{tabular}{c|c|cc|cc}
\toprule
\multirow[c]{2}{*}[-0.45ex]{\textbf{Model}}
& \multirow[c]{2}{*}[-0.45ex]{\textbf{Method}}
& \multicolumn{2}{c|}{\textbf{\ndonnx}}
& \multicolumn{2}{c}{\textbf{\numba}} \\
\cmidrule(lr){3-4}\cmidrule(lr){5-6}
& & \textbf{\pass{1}} & \textbf{\exec{1}}
  & \textbf{\pass{1}} & \textbf{\exec{1}} \\
\midrule

\multirow[c]{3}{*}[-0.35ex]{\shortstack{\fontfamily{pcr}\selectfont Qwen3\\\fontfamily{pcr}\selectfont-32B}}
& ReMem
& 36.69 & 41.42
& 54.55 & 72.19 \\
& DC-RS
& 21.30 & 23.08
& 40.11 & 57.75 \\
& \cellcolor[HTML]{FFF4D6}\name
& \cellcolor[HTML]{FFF4D6}\textbf{49.70}
& \cellcolor[HTML]{FFF4D6}\textbf{57.40}
& \cellcolor[HTML]{FFF4D6}\textbf{68.45}
& \cellcolor[HTML]{FFF4D6}\textbf{87.17} \\

\midrule

\multirow[c]{3}{*}[-0.35ex]{\shortstack{\fontfamily{pcr}\selectfont GLM-4\\\fontfamily{pcr}\selectfont-Plus}}
& ReMem
& 21.89 & 30.77
& 48.66 & 69.52 \\
& DC-RS
& 32.54 & 55.62
& 17.11 & 21.93 \\
& \cellcolor[HTML]{FFF4D6}\name
& \cellcolor[HTML]{FFF4D6}\textbf{52.66}
& \cellcolor[HTML]{FFF4D6}\textbf{60.36}
& \cellcolor[HTML]{FFF4D6}\textbf{71.12}
& \cellcolor[HTML]{FFF4D6}\textbf{92.51} \\

\bottomrule
\end{tabular}
}
\vspace{-0.15in}
\end{table}

\subsection{Implementation Details}
\label{sec:implementation}

\looseness=-1
We use \texttt{bge-base-en-v1.5} as the embedding model for all semantic retrieval modules. Unless otherwise specified, API documentation retrieval returns the top-5 APIs. For locally deployed models, we generate $n=10$ samples for each task with a temperature 0.7 and \texttt{top\_p} of 0.95, and report \pass{1}, \pass{3}, \pass{5}, \exec{1}, \exec{3}, and \exec{5}. For API-based large-model experiments, we set $n=1$ and temperature to 0 to reduce sampling randomness and API cost, and report \pass{1} and \exec{1}. All experiments use the same random seed.

\looseness=-1
For \name, memory retrieval is performed on top of the APIs retrieved by the corresponding RAG backbone. API-level and Cross-API-level memory each inject at most one successful code snippet and one usage guideline for each matched API or cross-API edge. For Task-level memory, we first retrieve historical tasks with semantic similarity above 0.8, rerank them by the utility-weighted similarity score defined in Section~\ref{sec:method-generation}, and inject the top-2 task memories. For self-evolving memory baselines, we adapt them to the same private-library setting by equipping them with Naive RAG for API documentation retrieval, and both ReMem and DC-RS retrieve the top-4 historical experiences for each new task.

\section{Experimental Results}
\label{sec:exp}

\subsection{RQ1: Effectiveness of \name in Private-Library Code Generation}
\label{sec:rq1}

\looseness=-1
In this RQ, we evaluate whether \name can effectively enhance existing RAG-based private-library code generation systems.

\noindent \textbf{Settings.}
We integrate \name with three representative RAG backbones, including \baselinename{Naive RAG}, \baselinename{EpiGEN}, and \baselinename{CAPIR}, and evaluate its effectiveness across LLMs of different scales.

\noindent \textbf{Results.}
The results for locally deployed models are reported in Table~\ref{tab:rag_comparison}, and the results for API-based larger models are reported in Table~\ref{tab:api_rag_comparison}.

\looseness=-1
\ding{182} \textit{\name consistently improves all evaluated RAG backbones across LLMs of different scales.}
As shown in Tables~\ref{tab:rag_comparison} and~\ref{tab:api_rag_comparison}, integrating \name brings consistent improvements over all RAG backbones on both \ndonnx and \numba. Across all evaluated settings, \name improves \pass{1} by 18.41 percentage points on average and \exec{1} by 21.17 percentage points on average. For locally deployed models, the gains are substantial across different model families. For example, with \texttt{Qwen2.5-Coder} on \ndonnx, \name improves the \pass{1} of \baselinename{Naive RAG}, \baselinename{EpiGEN}, and \baselinename{CAPIR} by 16.57, 21.30, and 19.88 percentage points, respectively. With \texttt{Llama-3.1} on \numba, \baselinename{Naive RAG}+\name further improves \pass{1} from 6.31\% to 45.99\% and \exec{1} from 23.58\% to 77.97\%. Similar improvements are also observed on stronger API-based models. For instance, with \texttt{GLM-4-Plus} on \ndonnx, \name improves \baselinename{CAPIR} from 33.14\% to 69.23\% in \pass{1}; with \texttt{Qwen3-32B}, it improves \baselinename{CAPIR} from 32.54\% to 62.72\%. These results show that \name is not tied to a specific model family or model scale, but provides generally useful private-library usage knowledge beyond static API documentation.

\looseness=-1
\ding{183} \textit{\name provides larger benefits than retrieval optimization alone.}
Comparing basic RAG enhanced with \name against standalone advanced RAG methods shows that self-evolving memory often brings stronger improvements than improving retrieval alone. For example, on \numba with \texttt{Llama-3.1}, \baselinename{Naive RAG}+\name achieves 45.99\% \pass{1}, substantially outperforming standalone \baselinename{EpiGEN} and \baselinename{CAPIR}, which achieve only 6.90\% and 5.13\%, respectively. On \ndonnx with \texttt{Qwen3-32B}, \baselinename{Naive RAG}+\name reaches 49.70\% \pass{1}, higher than standalone \baselinename{EpiGEN} and \baselinename{CAPIR}; with \texttt{GLM-4-Plus}, \baselinename{Naive RAG}+\name also outperforms standalone \baselinename{EpiGEN} and \baselinename{CAPIR} on \numba. These comparisons suggest that retrieval optimization mainly improves access to relevant API documents, whereas \name further supplies reusable \textit{how-to-use} knowledge distilled from execution feedback, which is critical for correctly invoking private-library APIs and composing them into valid solutions.

\looseness=-1
\ding{184} \textit{Case study: evolved memories complement static API documentation with practical usage constraints.}
Fig.~\ref{fig:rq3_memory_reuse_case} shows a representative case. Both Vanilla RAG and \name retrieve relevant documentation for \texttt{ndonnx.greater\_equal} and \texttt{ndonnx.divide}. However, Vanilla RAG still calls \texttt{ndonnx.greater\_equal} with an operand that is not an \texttt{ndonnx.Array}, leading to a type error. In contrast, \name retrieves an API-level memory learned from previous execution feedback, which records that raw inputs should first be converted with \texttt{ndonnx.asarray}. The generated code follows this guideline and passes the test. This case illustrates why \name can improve RAG-based generation: it reuses execution-derived API usage knowledge that is not explicit enough in static documentation.

\begin{boxK}
\small \textbf{Answer to RQ1:}
\name consistently enhances RAG-based private-library code generation across different retrieval backbones and LLM scales, demonstrating its effectiveness as a plug-in enhancement to existing RAG pipelines.
\end{boxK}

\subsection{RQ2: Effectiveness Compared to Self-evolving Memory Methods}
\label{sec:rq2}

\looseness=-1
In this RQ, we evaluate whether the multi-level evolving memory in \name is more effective for private-library code generation than general self-evolving memory methods.

\noindent \textbf{Settings.}
We compare \name with two representative memory-based baselines, \baselinename{ReMem} and \baselinename{DC-RS}. For a fair comparison, we equip all methods with the same Naive RAG module for retrieving static API documentation, and let them process tasks sequentially with execution feedback.

\noindent \textbf{Results.}
The results for locally deployed models are reported in Table~\ref{tab:cl_comparison}, and the results for API-based larger models are reported in Table~\ref{tab:api_cl_comparison}.

\looseness=-1
\ding{182} \textit{\name consistently outperforms existing self-evolving memory methods across different LLM scales.}
As shown in Tables~\ref{tab:cl_comparison} and~\ref{tab:api_cl_comparison}, \name achieves the best performance in all evaluated model-benchmark settings. Compared with the stronger baseline between \baselinename{ReMem} and \baselinename{DC-RS}, \name improves \pass{1} by 20.32 percentage points on average and \exec{1} by 22.76 percentage points on average. For locally deployed models, the gains are particularly large. On \numba with \texttt{Llama-3.1}, \name achieves 45.99\% \pass{1}, substantially outperforming \baselinename{ReMem} and \baselinename{DC-RS}, which achieve only 3.74\% and 8.24\%, respectively. On \ndonnx with \texttt{DeepSeek-Coder}, \name reaches 44.79\% \pass{1}, while \baselinename{ReMem} and \baselinename{DC-RS} reach 22.49\% and 15.74\%. The same trend also holds for stronger API-based models. For example, on \numba, \name improves \texttt{Qwen3-32B} to 68.45\% \pass{1}, compared with 54.55\% for \baselinename{ReMem} and 40.11\% for \baselinename{DC-RS}; with \texttt{GLM-4-Plus}, \name further achieves 71.12\% \pass{1}, clearly higher than both baselines.

\looseness=-1
\ding{183} \textit{Specialized multi-level memory provides more reliable adaptation than generic experience accumulation.}
The advantage of \name is especially clear in executability. On \numba with \texttt{Llama-3.1}, \name achieves 77.97\% \exec{1}, while \baselinename{ReMem} and \baselinename{DC-RS} only reach 22.51\% and 24.22\%, respectively. On \numba with \texttt{DeepSeek-Coder}, \name improves \exec{1} to 49.57\%, compared with 15.40\% for \baselinename{ReMem} and 14.65\% for \baselinename{DC-RS}. Larger models show similar patterns: on \numba, \name achieves 87.17\% \exec{1} with \texttt{Qwen3-32B} and 92.51\% with \texttt{GLM-4-Plus}. These results suggest that simply storing previous trajectories or synthesizing general cheatsheets is insufficient for private-library code generation. In contrast, \name explicitly organizes execution-derived knowledge into API-level, cross-API-level, and task-level memories, allowing it to capture API constraints, API coordination patterns, and reusable task-solving strategies more effectively.

\begin{boxK}
\small \textbf{Answer to RQ2:}
\name achieves stronger adaptation than existing self-evolving memory methods by combining multi-level evolving memory with feedback-driven evolution across API, cross-API, and task levels.
\end{boxK}

\subsection{RQ3: Ablation Study}
\label{sec:rq3}

\looseness=-1
In this RQ, we analyze the individual contributions of API-level, cross-API-level, and task-level memories in \name.

\noindent \textbf{Settings.}
We conduct ablation studies on \texttt{Llama-3.1} and \texttt{GLM-4-Plus}. We compare the full \name with three variants that disable API-level, cross-API-level, and task-level memory respectively. For each variant, the corresponding memory content, retrieval process, and feedback-driven reflection/update are removed, while all other components remain unchanged.

\noindent \textbf{Results.}
The results for \pass{1} and \exec{1} are reported in Table~\ref{tab:ablation}. Overall, removing any memory level reduces \pass{1} across all model-benchmark settings, confirming that all three levels contribute to functional correctness. Task-level memory is especially important for the locally deployed model: on \numba with \texttt{Llama-3.1}, removing $\mathcal{M}_{\text{Task}}$ drops \pass{1} from 45.99\% to 15.88\% and \exec{1} from 77.97\% to 38.50\%. This suggests that smaller models benefit strongly from reusable task-solving patterns accumulated from previous executions. API-level memory is crucial for reliable API invocation; for example, on \numba with \texttt{GLM-4-Plus}, removing $\mathcal{M}_{\text{API}}$ reduces \pass{1} from 71.12\% to 53.48\% and \exec{1} from 92.51\% to 71.66\%. Cross-API memory further improves API composition, as removing $\mathcal{M}_{\text{Cross}}$ on the same setting reduces \pass{1} to 54.55\% and \exec{1} to 74.33\%. These results show that the three memory levels capture complementary knowledge: individual API usage constraints, inter-API coordination patterns, and task-level solution strategies.

\begin{boxK}
\small \textbf{Answer to RQ3:}
All three memory levels are beneficial, validating the multi-level evolving memory design of \name.
\end{boxK}
\begin{table}[t]
\centering
\small
\renewcommand{\arraystretch}{1.1}
\setlength{\tabcolsep}{4pt}
\caption{Ablation study of \name in terms of \pass{1} and \exec{1} (\%) on \ndonnx and \numba.}
\vspace{-0.1in}
\label{tab:ablation}
\resizebox{\columnwidth}{!}{
\begin{tabular}{c|c|cc|cc}
\toprule
\multirow[c]{2}{*}[-0.45ex]{\textbf{Model}}
& \multirow[c]{2}{*}[-0.45ex]{\textbf{Setting}}
& \multicolumn{2}{c|}{\textbf{\ndonnx}}
& \multicolumn{2}{c}{\textbf{\numba}} \\
\cmidrule(lr){3-4}\cmidrule(lr){5-6}
& & \textbf{\pass{1}} & \textbf{\exec{1}}
& \textbf{\pass{1}} & \textbf{\exec{1}} \\
\midrule

\multirow[c]{4}{*}[-0.35ex]{\shortstack{\fontfamily{pcr}\selectfont Llama\\\fontfamily{pcr}\selectfont-3.1}}
& w/o $\mathcal{M}_{\text{API}}$
& 26.75 & 39.76
& 40.11 & 70.59 \\

& w/o $\mathcal{M}_{\text{Cross}}$
& 33.31 & 47.75
& 42.99 & 75.03 \\

& w/o $\mathcal{M}_{\text{Task}}$
& 20.65 & 37.04
& 15.88 & 38.50 \\

& \cellcolor[HTML]{FFF4D6}\textbf{Full \name}
& \cellcolor[HTML]{FFF4D6}\textbf{34.50}
& \cellcolor[HTML]{FFF4D6}47.63
& \cellcolor[HTML]{FFF4D6}\textbf{45.99}
& \cellcolor[HTML]{FFF4D6}\textbf{77.97} \\

\midrule

\multirow[c]{4}{*}[-0.35ex]{\shortstack{\fontfamily{pcr}\selectfont GLM-4\\\fontfamily{pcr}\selectfont-Plus}}
& w/o $\mathcal{M}_{\text{API}}$
& 41.42 & 51.48
& 53.48 & 71.66 \\

& w/o $\mathcal{M}_{\text{Cross}}$
& 47.34 & 56.21
& 54.55 & 74.33 \\

& w/o $\mathcal{M}_{\text{Task}}$
& 47.34 & 57.40
& 68.45 & 88.77 \\

& \cellcolor[HTML]{FFF4D6}\textbf{Full \name}
& \cellcolor[HTML]{FFF4D6}\textbf{52.66}
& \cellcolor[HTML]{FFF4D6}\textbf{60.36}
& \cellcolor[HTML]{FFF4D6}\textbf{71.12}
& \cellcolor[HTML]{FFF4D6}\textbf{92.51} \\

\bottomrule
\end{tabular}
}
\vspace{-0.15in}
\end{table}

\section{Discussion}
\label{sec:discussion}

\subsection{Threats to Validity}
\label{sec:discussion-threats}

\looseness=-1
\ding{182} \textit{Availability of Executable Feedback.}
\name relies on executable feedback to evolve its memory. In our experiments, this feedback is obtained from benchmark test cases, including pass/fail signals, runtime errors, and assertion failures. In practice, such feedback is often available in test-driven development workflows, where developers specify expected behaviors through unit tests before or during implementation. When manually written tests are incomplete, LLM-generated tests can also provide additional execution signals. For example, CodeT shows that generated tests can be used to evaluate and select candidate programs~\cite{chen2022codet}. This makes \name applicable to realistic development scenarios where executable checks can be obtained from manually written or LLM-assisted test cases.

\looseness=-1
\ding{183} \textit{Coverage of Private Libraries and Tasks.}
Evaluating private-library code generation is challenging because real enterprise private libraries are usually inaccessible to researchers. Following prior work, we use recently released public libraries to approximate private-library settings. Although \ndonnx and \numba are designed to be unfamiliar to evaluated models, they cannot fully represent all industrial private libraries. Moreover, public benchmarks constrain model selection, since the latest models may have been exposed to the benchmark libraries or related data. To reduce this threat, we evaluate \name across models with different scales and deployment types, and with multiple retrieval backends, including Naive RAG, EpiGEN, and CAPIR. Broader validation on industrial private-library repositories remains future work.

\looseness=-1
\ding{184} \textit{Dependence on LLM Reflection Quality.}
\name relies on LLMs to interpret execution feedback and update memories, so inaccurate reflections may introduce noisy guidance. To examine this risk, we manually sample 150 memory-evolution records from \texttt{Qwen3-32B}, with 30 samples for each inspected setting. We evaluate two aspects: error attribution, i.e., whether the LLM correctly judges whether the inspected API usage or API-composition usage is erroneous under the observed feedback, and guideline correctness, i.e., whether the generated memory update is relevant to the inspected target, factually correct, and practically useful for later generation. The inspected results show 87\% and 80\% accuracy for API-level and Cross-API-level error attribution, and 87\%, 80\%, and 70\% accuracy for API-level, Cross-API-level, and Task-level guideline correctness, respectively. Overall, the average accuracy across the 150 inspected records is 80.8\%, suggesting that the reflection process is generally reliable.

\looseness=-1
\ding{185} \textit{Long-term Memory Accumulation.}
Memory-based methods generally face memory growth as more tasks are processed. In \name, API-level and Cross-API-level memories are naturally bounded, since each API or cross-API edge keeps only one successful code memory and one current usage guideline. Task-level memory grows with the task stream, but moderate accumulation is useful for reusing task-solving experience. To prevent redundant or low-quality task memories from occupying storage and affecting later generation, \name introduces a monitoring layer that records usage frequency and historical success/failure outcomes. This provides a practical mechanism to control memory accumulation by pruning rarely used or low-utility task memories when the memory becomes too large.

\begin{figure}[t]
    \centering
    \setlength{\abovecaptionskip}{0pt}
    \setlength{\belowcaptionskip}{0pt}
    \includegraphics[width=\columnwidth]{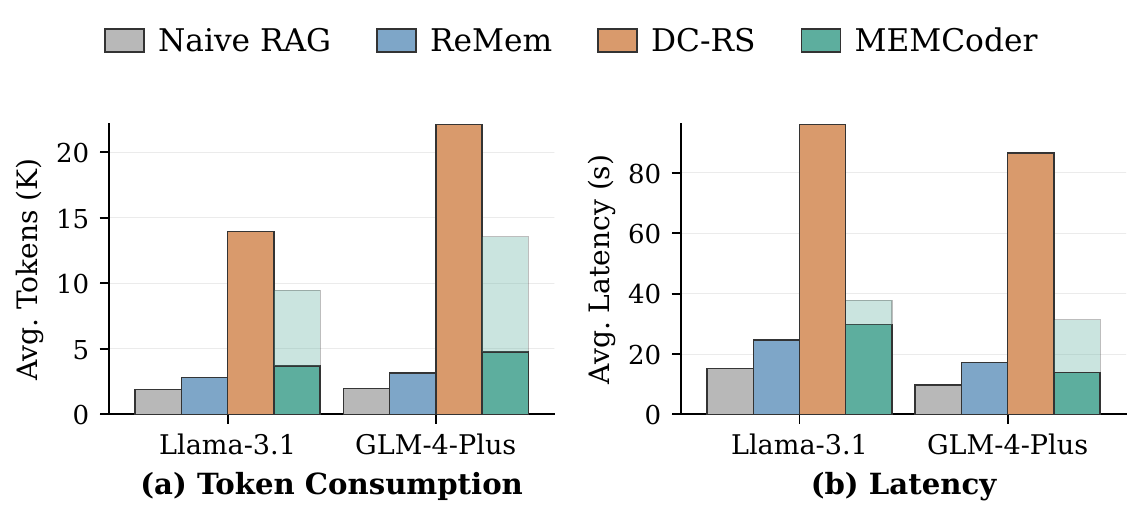}
    \caption{Average token usage and latency on \ndonnx and \numba. Solid and light segments indicate inference and memory evolution costs.}
    \label{fig:efficiency}
    \vspace{-14pt}
\end{figure}

\subsection{Computational Overhead and Inference Efficiency}
\label{sec:discussion-efficiency}

\looseness=-1
Although \name is training-free and does not update model parameters, its self-evolving memory introduces extra computation from memory retrieval, prompt injection, and feedback-driven memory evolution. We therefore analyze whether the correctness gains come with acceptable efficiency cost. To cover different model scales, we compare Naive RAG, ReMem, DC-RS, and \name on Llama-3.1-8B-Instruct and GLM-4-Plus.

\looseness=-1
As shown in Figure~\ref{fig:efficiency}, the inference-stage token cost of \name remains moderate. On Llama-3.1 and GLM-4-Plus, \name consumes 3.71K and 4.78K inference tokens on average, only slightly higher than Naive RAG and ReMem because of memory injection. This overhead is much smaller than DC-RS, which consumes 13.97K and 22.16K tokens due to its longer generated experiences. Meanwhile, Tables~\ref{tab:cl_comparison} and~\ref{tab:api_cl_comparison} show that \name achieves higher \pass{1} and \exec{1} than ReMem and DC-RS in most settings, suggesting that the additional tokens are effectively used. Even when memory evolution is counted, the monetary cost remains small: under the GLM-4-Plus price of 5 RMB per million tokens, \name adds about 4.09M tokens over Naive RAG on the two benchmarks, corresponding to approximately \$2.84.

\looseness=-1
The latency overhead is also acceptable. Since memory evolution occurs after execution feedback is obtained, it can be performed asynchronously and does not need to block the current response. Therefore, user-facing latency is mainly determined by inference latency, shown as the solid segments in Figure~\ref{fig:efficiency}. The inference latency of \name is 29.74s on Llama-3.1 and 13.92s on GLM-4-Plus, close to ReMem and substantially lower than DC-RS. Overall, \name introduces moderate token and latency overhead while providing substantial correctness improvements.

\section{Conclusion}
\label{sec:conclusion}

\looseness=-1
In this paper, we study \textit{Private-Library Code Generation}, where LLMs must use APIs absent from public pre-training corpora. Our analysis shows that static API documentation, even when retrieved by an oracle, mainly helps identify \textit{what APIs to use} but cannot sufficiently teach \textit{how to use them correctly}, leading to recurring errors at the API, cross-API, and task levels. To address this, we propose \name, a training-free self-evolving memory framework that augments RAG pipelines with a \textbf{Multi-level Evolving Memory}. \name accumulates execution-derived \textit{Usage Guidelines}, retrieves them during generation, and refines them after execution feedback. Experiments on \ndonnx and \numba show that \name consistently enhances different RAG backbones, outperforms existing self-evolving memory methods, and validates the contribution of multi-level usage memories. These results demonstrate that self-evolving usage memory is a practical plug-in enhancement for RAG-based private-library code generation.

\section{DATA AVAILABILITY}
\label{sec:avalability}
The artifacts are publicly available in the GitHub repository at https://github.com/THU-Agent/MEMCoder.git.



\bibliographystyle{ieeetr}
\bibliography{doc}

@misc{li2024revisiting,
      title={Revisiting Catastrophic Forgetting in Large Language Model Tuning}, 
      author={Hongyu Li and Liang Ding and Meng Fang and Dacheng Tao},
      year={2024},
      eprint={2406.04836},
      archivePrefix={arXiv},
      primaryClass={cs.CL},
      url={https://arxiv.org/abs/2406.04836}, 
}

@misc{luo2025empirical,
      title={An Empirical Study of Catastrophic Forgetting in Large Language Models During Continual Fine-tuning}, 
      author={Yun Luo and Zhen Yang and Fandong Meng and Yafu Li and Jie Zhou and Yue Zhang},
      year={2025},
      eprint={2308.08747},
      archivePrefix={arXiv},
      primaryClass={cs.CL},
      url={https://arxiv.org/abs/2308.08747}, 
}

@misc{zhang2026masterteachingllmsuse,
      title={To See is Not to Master: Teaching LLMs to Use Private Libraries for Code Generation}, 
      author={Yitong Zhang and Chengze Li and Ruize Chen and Guowei Yang and Xiaoran Jia and Yijie Ren and Jia Li},
      year={2026},
      eprint={2603.15159},
      archivePrefix={arXiv},
      primaryClass={cs.SE},
      url={https://arxiv.org/abs/2603.15159}, 
}

@article{hui2024qwen25coder,
  author       = {Binyuan Hui and
                  Jian Yang and
                  Zeyu Cui and
                  Jiaxi Yang and
                  Dayiheng Liu and
                  Lei Zhang and
                  Tianyu Liu and
                  Jiajun Zhang and
                  Bowen Yu and
                  Kai Dang and
                  An Yang and
                  Rui Men and
                  Fei Huang and
                  Xingzhang Ren and
                  Xuancheng Ren and
                  Jingren Zhou and
                  Junyang Lin},
  title        = {Qwen2.5-Coder Technical Report},
  journal      = {CoRR},
  volume       = {abs/2409.12186},
  year         = {2024}
}

@misc{suzgun2025dynamic,
      title={Dynamic Cheatsheet: Test-Time Learning with Adaptive Memory}, 
      author={Mirac Suzgun and Mert Yuksekgonul and Federico Bianchi and Dan Jurafsky and James Zou},
      year={2025},
      eprint={2504.07952},
      archivePrefix={arXiv},
      primaryClass={cs.LG},
      url={https://arxiv.org/abs/2504.07952}, 
}

@misc{wei2025evomem,
      title={Evo-Memory: Benchmarking LLM Agent Test-time Learning with Self-Evolving Memory}, 
      author={Tianxin Wei and Noveen Sachdeva and Benjamin Coleman and Zhankui He and Yuanchen Bei and Xuying Ning and Mengting Ai and Yunzhe Li and Jingrui He and Ed H. Chi and Chi Wang and Shuo Chen and Fernando Pereira and Wang-Cheng Kang and Derek Zhiyuan Cheng},
      year={2025},
      eprint={2511.20857},
      archivePrefix={arXiv},
      primaryClass={cs.CL},
      url={https://arxiv.org/abs/2511.20857}, 
}

@article{li2025structured,
  title={Structured chain-of-thought prompting for code generation},
  author={Li, Jia and Li, Ge and Li, Yongmin and Jin, Zhi},
  journal={ACM Transactions on Software Engineering and Methodology},
  volume={34},
  number={2},
  pages={1--23},
  year={2025},
  publisher={ACM New York, NY}
}

@inproceedings{jiang2025aixcoder,
  title={aiXcoder-7B: A Lightweight and Effective Large Language Model for Code Processing},
  author={Jiang, Siyuan and Li, Jia and Zong, He and Liu, Huanyu and Zhu, Hao and Hu, Shukai and Li, Erlu and Ding, Jiazheng and Han, Yu and Ning, Wei and others},
  booktitle={2025 IEEE/ACM 47th International Conference on Software Engineering: Software Engineering in Practice (ICSE-SEIP)},
  pages={215--226},
  year={2025},
  organization={IEEE}
}

@article{cai2025ai,
  title={AI-Driven Self-Evolving Software: A Promising Path Toward Software Automation},
  author={Cai, Liyi and Ren, Yijie and Zhang, Yitong and Li, Jia},
  journal={arXiv preprint arXiv:2510.00591},
  year={2025}
}

@article{li2025beyond,
  title={Beyond autoregression: An empirical study of diffusion large language models for code generation},
  author={Li, Chengze and Zhang, Yitong and Li, Jia and Cai, Liyi and Li, Ge},
  journal={arXiv preprint arXiv:2509.11252},
  year={2025}
}

@inproceedings{apifinder,
  title={When language model meets private library},
  author={Zan, Daoguang and Chen, Bei and Lin, Zeqi and Guan, Bei and Yongji, Wang and Lou, Jian-Guang},
  booktitle={Findings of the Association for Computational Linguistics: EMNLP 2022},
  pages={277--288},
  year={2022}
}

@inproceedings{exploracoder,
  title={ExploraCoder: Advancing code generation for multiple unseen APIs via planning and chained exploration},
  author={Wang, Yunkun and Zhang, Yue and Qin, Zhen and Zhi, Chen and Li, Binhua and Huang, Fei and Li, Yongbin and Deng, Shuiguang},
  booktitle={Proceedings of the 63rd Annual Meeting of the Association for Computational Linguistics (Volume 1: Long Papers)},
  pages={18124--18145},
  year={2025}
}

@inproceedings{docprompting,
  title={Docprompting: Generating code by retrieving the docs},
  author={Zhou, Shuyan and Alon, Uri and Xu, Frank F and Jiang, Zhengbao and Neubig, Graham},
  booktitle={The Eleventh International Conference on Learning Representations},
  year={2022}
}

@inproceedings{epigen,
  title={Epigen: An efficient multi-api code generation framework under enterprise scenario},
  author={Li, Sijie and Li, Sha and Zhang, Hao and Li, Shuyang and Chen, Kai and Yuan, Jianyong and Cao, Yi and Yang, Lvqing},
  booktitle={Proceedings of the 2024 Joint International Conference on Computational Linguistics, Language Resources and Evaluation (LREC-COLING 2024)},
  pages={6206--6215},
  year={2024}
}

@article{llama,
  title={The llama 3 herd of models},
  author={Grattafiori, Aaron and Dubey, Abhimanyu and Jauhri, Abhinav and Pandey, Abhinav and Kadian, Abhishek and Al-Dahle, Ahmad and Letman, Aiesha and Mathur, Akhil and Schelten, Alan and Vaughan, Alex and others},
  journal={arXiv preprint arXiv:2407.21783},
  year={2024}
}

@inproceedings{capir,
  title={Compositional API recommendation for library-oriented code generation},
  author={Ma, Zexiong and An, Shengnan and Xie, Bing and Lin, Zeqi},
  booktitle={Proceedings of the 32nd IEEE/ACM International Conference on Program Comprehension},
  pages={87--98},
  year={2024}
}

@misc{ndonnx,
  title        = {ndonnx (Version 0.17.1)},
  author       = {QuantCo},
  year         = {2025},
  howpublished = {\url{https://pypi.org/project/ndonnx/0.17.1/}},
}

@article{yang2025qwen3,
  title={Qwen3 technical report},
  author={Yang, An and Li, Anfeng and Yang, Baosong and Zhang, Beichen and Hui, Binyuan and Zheng, Bo and Yu, Bowen and Gao, Chang and Huang, Chengen and Lv, Chenxu and others},
  journal={arXiv preprint arXiv:2505.09388},
  year={2025}
}

@article{deepseekcoder,
  title={DeepSeek-Coder: when the large language model meets programming--the rise of code intelligence},
  author={Guo, Daya and Zhu, Qihao and Yang, Dejian and Xie, Zhenda and Dong, Kai and Zhang, Wentao and Chen, Guanting and Bi, Xiao and Wu, Yifan and Li, YK and others},
  journal={arXiv preprint arXiv:2401.14196},
  year={2024}
}

@inproceedings{liu2023codegen4libs,
  title={Codegen4libs: A two-stage approach for library-oriented code generation},
  author={Liu, Mingwei and Yang, Tianyong and Lou, Yiling and Du, Xueying and Wang, Ying and Peng, Xin},
  booktitle={2023 38th IEEE/ACM International Conference on Automated Software Engineering (ASE)},
  pages={434--445},
  year={2023},
  organization={IEEE}
}

@inproceedings{liu2025think,
  title={THINK: Tackling API Hallucinations in LLMs via Injecting Knowledge},
  author={Liu, Jiaxin and Zhang, Yating and Wang, Deze and Li, Yiwei and Dong, Wei},
  booktitle={2025 IEEE International Conference on Software Analysis, Evolution and Reengineering (SANER)},
  pages={229--240},
  year={2025},
  organization={IEEE}
}

@article{gu2025effectiveness,
  title={On the effectiveness of large language models in domain-specific code generation},
  author={Gu, Xiaodong and Chen, Meng and Lin, Yalan and Hu, Yuhan and Zhang, Hongyu and Wan, Chengcheng and Wei, Zhao and Xu, Yong and Wang, Juhong},
  journal={ACM Transactions on Software Engineering and Methodology},
  volume={34},
  number={3},
  pages={1--22},
  year={2025},
  publisher={ACM New York, NY}
}

@article{zan2024diffcoder,
  title={DiffCoder: Enhancing large language model on API invocation via analogical code exercises},
  author={Zan, Daoguang and Yu, Ailun and Shen, Bo and Chen, Bei and Li, Wei and Gong, Yongshun and Chen, Xiaolin and Yao, Yafen and Luo, Weihua and Guan, Bei and others},
  journal={Proceedings of the ACM on Software Engineering},
  volume={1},
  number={FSE},
  pages={406--426},
  year={2024},
  publisher={ACM New York, NY, USA}
}

@article{lewis2020retrieval,
  title={Retrieval-augmented generation for knowledge-intensive nlp tasks},
  author={Lewis, Patrick and Perez, Ethan and Piktus, Aleksandra and Petroni, Fabio and Karpukhin, Vladimir and Goyal, Naman and K{\"u}ttler, Heinrich and Lewis, Mike and Yih, Wen-tau and Rockt{\"a}schel, Tim and others},
  journal={Advances in neural information processing systems},
  volume={33},
  pages={9459--9474},
  year={2020}
}

@article{zan2025private,
  title={Private-library-oriented code generation with large language models},
  author={Zan, Daoguang and Chen, Bei and Gong, Yongshun and Cao, Junzhi and Zhang, Fengji and Wu, Bingchao and Guan, Bei and Yin, Yilong and Wang, Yongji},
  journal={Knowledge-Based Systems},
  volume={326},
  pages={113934},
  year={2025},
  publisher={Elsevier}
}

@article{chen2022codet,
  title={Codet: Code generation with generated tests},
  author={Chen, Bei and Zhang, Fengji and Nguyen, Anh and Zan, Daoguang and Lin, Zeqi and Lou, Jian-Guang and Chen, Weizhu},
  journal={arXiv preprint arXiv:2207.10397},
  year={2022}
}

@article{athiwaratkun2022multi,
  title={Multi-lingual evaluation of code generation models},
  author={Athiwaratkun, Ben and Gouda, Sanjay Krishna and Wang, Zijian and Li, Xiaopeng and Tian, Yuchen and Tan, Ming and Ahmad, Wasi Uddin and Wang, Shiqi and Sun, Qing and Shang, Mingyue and others},
  journal={arXiv preprint arXiv:2210.14868},
  year={2022}
}

@article{glm2024chatglm,
  title={Chatglm: A family of large language models from glm-130b to glm-4 all tools},
  author={Glm, Team and Zeng, Aohan and Xu, Bin and Wang, Bowen and Zhang, Chenhui and Yin, Da and Zhang, Dan and Rojas, Diego and Feng, Guanyu and Zhao, Hanlin and others},
  journal={arXiv preprint arXiv:2406.12793},
  year={2024}
}

\end{document}